\documentclass[conference]{IEEEtran}

\usepackage[nolist]{acronym}
\usepackage[T1]{fontenc}
\usepackage{calc}
\usepackage{multirow}
\usepackage{tikz}
\usetikzlibrary[shapes.geometric]
\usetikzlibrary{shapes.misc}
\usetikzlibrary{calc}
\usetikzlibrary{backgrounds}
\usetikzlibrary{decorations.pathreplacing,angles,quotes}
\usetikzlibrary{shadings}
\usepackage{makecell}
\usepackage{algorithm}
\usepackage{algpseudocode}
\algrenewcommand\algorithmicindent{0.8em}%
\algrenewcommand\algorithmiccomment[1]{\hfill\textcolor{gray}{// #1}}

\usepackage{amsfonts}
\usepackage{amssymb}
\usepackage{pgfkeys}
\usepackage{collcell}

\usepackage{booktabs}
\usepackage{harveyballs}
\usepackage{collcell}

\usepackage{xcolor}
\PassOptionsToPackage{table}{xcolor}

\usepackage{colortbl}
\usepackage{array,tabularx}
\usepackage{adjustbox}
\usepackage{diagbox}

\usepackage[most]{tcolorbox}

\usepackage{adjustbox}
\usepackage{caption}
\usepackage{subcaption}
\usepackage{pgfplots}
\usepackage{pgfplotstable}

\usepackage{soul}
\usepackage{tabularx}

\pgfplotsset{compat=1.18} 
\usepackage{circledtext}
\usepackage{pifont}
\usepackage{fontawesome5}

\usepackage{url}
\usepackage{hyperref}
\hypersetup{hidelinks}
\usepackage{microtype}

\usepgfplotslibrary{groupplots}
\pgfplotsset{compat=newest}
\usepackage{longtable}
\usepackage{stackengine}

\usepackage{pifont}
\newcolumntype{L}[1]{>{\raggedright\let\newline\\\arraybackslash\hspace{0pt}}m{#1}}
\newcolumntype{C}[1]{>{\centering\let\newline\\\arraybackslash\hspace{0pt}}m{#1}}
\usepackage[inline]{enumitem}

\renewcommand{\subsubsection}[1]{{\vspace*{0.2\baselineskip}\noindent\bfseries #1:\quad}}

\newcommand{\etal}{~\textit{et al.}}
\newcommand{\povd}{\ac{POV} display}

\newcommand{\circledRomanText}[1]{%
  \raisebox{.5pt}{\textcircled{\raisebox{-.2pt}{\centering\scriptsize\uppercase\expandafter{\romannumeral #1\relax}}}}%
}

\definecolor{ir}{rgb}{0.9294,0.8549,0.9647}

\def\fullcircle#1{
    \tikz[]{
        \node[circle, fill=ir, draw=ir!50!black, minimum size=#1] at (0, 0) {};
    }
}

\def\twoslices#1{
    \tikz[]{
        \filldraw[fill=ir, draw=ir!50!black] (0, 0) -- ++(#1,0) arc[start angle=0, end angle=120, radius=#1] -- cycle;
        \begin{scope}[cm={-1,0,0,-1,(0, 0)}]
        \filldraw[fill=ir, draw=ir!50!black] (0, 0) -- ++(#1,0) arc[start angle=0, end angle=120, radius=#1] -- cycle;
        \end{scope}
    }
}
\def\fullcircleRGB#1#2{
    \tikz[]{
        \node[
            circle,
            fill=#2!50,
            draw=#2!50!black,
            minimum size=#1
        ] at (0, 0) {};
    }
}

\newcommand{\lightbulb}[1]{%
  \ifnum\pdfstrcmp{#1}{yellow}=0
    \includegraphics[height=1em]{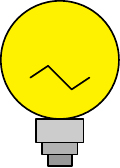}%
  \else\ifnum\pdfstrcmp{#1}{black}=0
    \includegraphics[height=1em]{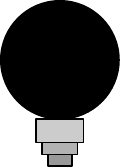}%
  \else\ifnum\pdfstrcmp{#1}{yellow!40!black}=0
    \includegraphics[height=1em]{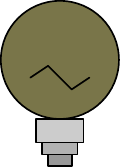}%
  \fi\fi\fi
}

\newcommand{\cmark}{\ding{51}}
\newcommand{\xmark}{\ding{55}}

\newcommand{\cmarkUS}{\cellcolor{blue!10}\ding{51}}

\IEEEoverridecommandlockouts

\begin{document}

\title{The Spectrum Strikes Back: Infrared POV Attacks on Traffic Sign Classification}

\author{\IEEEauthorblockN{Michael K\"uhr$^*$,
		Mevlüt Yildirim$^*$,
		Maximilian Luedecke,
		Mohammad Hamad,
		Sebastian Steinhorst}
	\IEEEauthorblockA{TUM School of Computation, Information and Technology, Technical University of Munich, Germany\\\{firstname.lastname\}@tum.de}
	\thanks{$^*$Both authors contributed equally to this work.}}

\maketitle

\begin{abstract}
Traffic sign classification is a crucial task for autonomous vehicles, and numerous attacks against it have been identified. A majority of physical adversarial attacks involve attaching patches to traffic signs or projecting perturbations on them. While they demonstrate high effectiveness, they are perceptible to humans. At the same time, light-based attacks outside the human visible spectrum are known but have limitations in their dynamic adaptability. We propose a persistence-of-vision-based attack that operates in the near-infrared light spectrum. With the possibility of showing dynamic, remotely triggered content, this allows a stealthy physical adversarial attack against traffic sign classification. By identifying the optimal position through digital simulation, we conduct extensive real-world evaluations using two different traffic signs, 12 machine learning models from different families, multiple distances up to 20 meters, and varying illumination conditions. Our evaluation shows high attack success rates across our test scenarios. We propose near-infrared cutoff filters and a software-based detection mechanism as defenses, and tackle limitations of the near-infrared persistence of vision display by prototyping a human-visible RGB version of it.
\end{abstract}

\IEEEpeerreviewmaketitle

\begin{acronym}
\acro{CMOS}{Complementary Metal-Oxide-Semiconductor}
\acro{POV}{persistence of vision}
\acro{LED}{light-emitting diode}
\acro{ML}{Machine Learning}
\acro{PWM}{pulse-width modulation}
\acro{PCB}{printed circuit board}
\acro{ASR}{attack success rate}
\acro{GTSRB}{German Traffic Sign Recognition Benchmark}
\acro{TN}{true negative}
\acro{TP}{true positive}
\acro{BFS}{breadth-first search}
\end{acronym}
\section{Introduction}

While autonomous vehicles are becoming reality~\cite{mercedes-benz_group_ag_mercedes-benz_2023, waymo_llc_self-driving_2024}, the attack surface of such cars increases with the integration of perception sensors~\cite{yan_sok_2020, gao_autonomous_2022, el-rewini_cybersecurity_2020}, such as LiDARs or cameras. Among these sensors, cameras are often used as a single source of information for tasks like traffic sign detection, lane detection, or traffic light recognition~\cite{gao_autonomous_2022, ibanez-guzman_lidar_2025}. As a result, camera-based perception has become a critical target for attacks. Specifically, physical adversarial attacks have been investigated by different research communities~\cite{guesmi_physical_2023, wei_physical_2024, badjie_adversarial_2025, kuhr_sok_2026}.

\begin{figure}[t]
    \centering
    \includegraphics[width=0.9\linewidth]{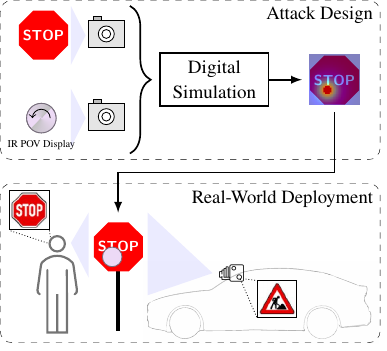}
    \caption{Infrared-based persistence of vision displays are a physical adversarial attack that can target traffic sign classification while being stealthy to humans. We identify the optimal placement through a digital simulation to ensure high attack success. As a result, traffic sign classifiers will misclassify the sign.}
    \label{fig:fig1}
\end{figure}

These attacks against camera-based perception have been explored using a wide range of techniques, including printed stickers on traffic signs~\cite{eykholt_robust_2018}, projector-based attacks~\cite{lovisotto_slap_2021}, or light-based attacks in the human-visible~\cite{duan_adversarial_2021} and invisible spectrum~\cite{sato_invisible_2024, wang_i_2021}. Such attacks target the \ac{ML} models that are used in autonomous vehicles to perform tasks such as traffic sign detection. Among them, attacks operating outside the human-visible spectrum are particularly concerning, as they show high stealthiness and are difficult or impossible to detect by humans~\cite{wang_i_2021}. 

Previous studies identified that autonomous vehicles can perceive light sources in this infrared spectrum~\cite{sato_invisible_2024, wang_i_2021}. Additionally, infrared-sensitive cameras are already used in driver assistance systems to increase perception in nighttime scenarios~\cite{aumovio_se_night-capable_2025} and are being researched by the automotive industry for use in adverse environmental conditions~\cite{dubbert_all-weather_2019}. This makes near-infrared attacks both practical and safety-critical. However, existing infrared-based physical attacks face important limitations: They rely on static perturbations or continuous light emission~\cite{wang_i_2021} and often require high-power laser beams~\cite{sato_invisible_2024}. As a result, these attacks lack flexibility, do not support selective triggering, and offer limited adaptability once deployed, which restricts their practicality in real-world settings.

Motivated by the limitations of existing physical adversarial attacks in \autoref{tab:SOTA_Table}, we introduce a new physical adversarial attack based on a \textit{dynamically reconfigurable and triggerable infrared light source}. As shown in \autoref{fig:fig1}, we realize this attack using a \povd{} that operates in the near-infrared spectrum and can display dynamic, remotely triggered adversarial patterns. Unlike prior infrared attacks, our \povd{} does not rely on high-power lasers~\cite{sato_invisible_2024} but instead operates using commercially available infrared \acp{LED}~\cite{kingbright_wp7113sf6bt-p22_2024}. By leveraging fast rotation, we exploit the \ac{POV} mechanism, and our \povd{} shows spatially structured infrared patterns without relying on static physical patches. This design enables three key capabilities that distinguish our attack from prior work
\begin{enumerate*}[before=\unskip{: }, itemjoin={{; }}, itemjoin*={{, and }}, label={(\roman*)}]
    \item the attack can display dynamically reconfigurable content, allowing adaptation to different target assets and models without redeployment
    \item it can be selectively activated by controlling the infrared \acp{LED} while maintaining rotation, enabling the adversarial effect without permanently altering the appearance of the traffic sign
    \item it remains highly stealthy, as the emitted infrared light is invisible to human observers and the rotating \povd{} appears transparent when inactive. 
\end{enumerate*}
These properties differentiate our approach from static physical attacks such as stickers~\cite{eykholt_robust_2018} and from human-visible projections~\cite{lovisotto_slap_2021}.

\begin{table}[t]
	\centering
	\caption{Comparison of existing physical attacks with our \povd{} attack.}
	\label{tab:SOTA_Table}
    \setlength{\tabcolsep}{2pt}
	\resizebox{1\linewidth}{!}{%
    \begin{tabular}{lC{0.15\linewidth}C{0.17\linewidth}C{0.15\linewidth}C{0.15\linewidth}C{0.18\linewidth}}
        \toprule
        \textbf{Attack} & \textbf{Dynamic} & \textbf{Triggered} & \textbf{Stealth} & \textbf{Remote} & \textbf{Transferable}\\ \bottomrule
        RP\textsubscript{2}~\cite{eykholt_robust_2018} & \xmark & \xmark & \xmark & \cmark & \xmark \\
        SLAP~\cite{lovisotto_slap_2021} & \cmark & \cmark & \xmark & \cmark & \cmark \\
        AdvLB~\cite{duan_adversarial_2021} & \xmark & \cmark & \xmark & \cmark & \xmark \\
        ILR attack~\cite{sato_invisible_2024} & \xmark & \cmark & \cmark & \cmark & \cmark \\
        ICSL Attack~\cite{wang_i_2021} & \xmark & \cmark & \cmark & \cmark & \xmark\\
        \cellcolor{blue!10}\textbf{Ours (\povd{})} & \cmarkUS & \cmarkUS & \cmarkUS & \cmarkUS & \cmarkUS \\
        \bottomrule
    \end{tabular}}
\end{table}

To deploy this attack reliably, we provide a digital simulation to identify the optimal placement of freely configurable content in the \povd{} on a traffic sign. By applying different transformations~\cite{athalye_synthesizing_2018} within our digital simulation, we ensure that our attack shows high robustness under different physical conditions, such as varying distances and illumination conditions. This simulation-driven placement optimization is critical for real-world effectiveness. Guided by the simulation results, we conduct extensive real-world evaluations across different traffic signs, \povd{} sizes, illumination conditions, and 12 \ac{ML} models spanning multiple architectures and training datasets, demonstrating high \aclp{ASR}. To mitigate this threat, we evaluate both a hardware-based defense using near-infrared cutoff filters and a software-based detection mechanism that exploits sensor-specific spectral artifacts. Finally, we discuss the broader implications of POV-based attacks and show that similar threats can also arise when deploying \povd{}s in the human-visible spectrum.

In summary, our contributions are:
\begin{itemize}[nosep]
    \item \textbf{Reconfigurable and triggerable infrared attack:}  We present a physical adversarial infrared \povd{} attack that enables high stealthiness and dynamically reconfigurable, remotely triggerable perturbations against camera-based perception. 
    \item \textbf{Digital simulation of our \povd{} attack:} We digitally simulate of our attack to identify the optimal positioning of the \povd.
    \item \textbf{Extensive real-world evaluation:} By performing physical tests with varying distances, displayed content, \povd{} sizes, and \ac{ML} models, we comprehensively evaluate our attack and demonstrate its effectiveness.
    \item \textbf{Defense against our attack, including evaluation:} We propose defense and detection methods to protect traffic sign classification models against our \povd{} attack.
\end{itemize}

\section{Background}

This section introduces background information on infrared light and the \ac{POV} effect on which our attack is based.

\subsection{Infrared Characteristics of Cameras}
Typical \ac{CMOS} image sensors that are used in camera-based perception pipelines in autonomous vehicles are sensitive to near-infrared light with a wavelength of 780nm to 1000nm~\cite{international_commission_on_illumination_cie_2016, gouveia_advances_2016, el_gamal_cmos_2005}. This wavelength is already outside the human-visible spectrum, which goes up to approximately 760nm~\cite{international_commission_on_illumination_cie_2016}, making it not perceivable by humans. At the same time, different attacks against camera-based perception are known that exploit the sensitivity of image sensors to infrared light, not only in autonomous vehicles~\cite{sato_invisible_2024,wang_i_2021}, but also in face detection~\cite{zhou_invisible_2018, wang_invisible_2024}. Based on the sensor-specific spectral sensitivity, near-infrared light is perceived either as red, purple, or magenta~\cite{sato_invisible_2024, wang_i_2021}.

\subsection{Persistence of Vision}
The so-called \ac{POV} effect tricks the human visual perception system~\cite{coltheart_persistences_1980, ferry_persistence_1892} and is colloquially referred to as "holograms" when used in \povd{}s. These \povd{}s typically consist of a rotating fan, equipped with precisely controlled \acp{LED}. At high rotation speeds, they trick the human visual information-processing system into a circle-shaped display. Such \povd{}s in the human-visible light spectrum are typically found for advertising purposes or at show events.

A similar effect is also observable with cameras: If the exposure time $t_{exp}$ is larger than or equal to the rotational speed $f_{rot}$ of the \povd{}, the complete displayed content will be captured. With smaller exposure times, only fractions of the displayed content are available. This leads to the exposure time constraint of \autoref{eq:povConstraint}.

\begin{equation}
    \label{eq:povConstraint}
    t_{exp}\geq\frac{1}{f_{rot}}
\end{equation}

\begin{figure}[t]
     \centering
     \begin{subfigure}[b]{0.23\linewidth}
         \centering
         \includegraphics[width=1\textwidth]{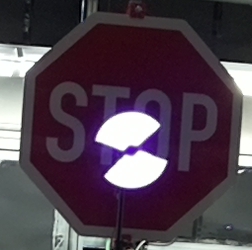}
         \caption{$t_{exp}<\frac{1}{f_{rot}}$}
         \label{fig:pov_effect_texp_smaller_frot}
     \end{subfigure}
     \hfill
     \begin{subfigure}[b]{0.23\linewidth}
         \centering
         \includegraphics[width=1\textwidth]{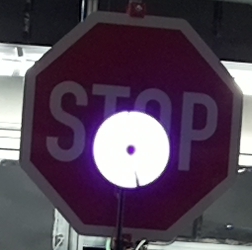}
         \caption{$t_{exp}\approx\frac{1}{f_{rot}}$}
         \label{fig:pov_effect_texp_equal_frot}
     \end{subfigure}
     \hfill
     \begin{subfigure}[b]{0.23\linewidth}
         \centering
         \includegraphics[width=1\textwidth]{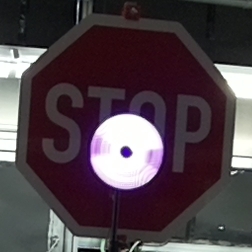}
         \caption{$t_{exp}>\frac{1}{f_{rot}}$}
         \label{fig:pov_effect_texp_greater_frot}
     \end{subfigure}
     \caption{A two-blade \povd{} in front of a stop sign, visualizing the exposure time constraint when showing a full circle. If $t_{exp}<\frac{1}{f_{rot}}$, the pattern cannot be captured completely, if $t_{exp}=\frac{1}{f_{rot}}$, the pattern is captured fully, and if $t_{exp}>\frac{1}{f_{rot}}$, artifacts such as brighter areas due to overlaps can occur.}
     \label{fig:pov_effect}
\end{figure}

If this time constraint is fulfilled and the \povd{} shows static content, artifacts coming from the rolling-shutter effect~\cite{el_gamal_cmos_2005} of \ac{CMOS} image sensors can be mostly neglected, since the complete displayed content will be captured. \autoref{fig:pov_effect} shows the effect of the time constraint from \autoref{eq:povConstraint} with different scenarios.

\section{Threat Model}

In this section, we explain our attack goal and requirements, as well as the necessary knowledge and capabilities of the attacker.

\subsection{Attack Goal and Requirements}
\label{sec:attackGoalRequirements}

The goal of our attack are misclassifications of \ac{ML}-based image classifiers, specifically traffic sign classification models, although we will also test with generically trained image classification models. Following the terminology of Zhu\etal{}~\cite{zhu_tpatch_2023}, we will consider an \textbf{Altering Attack}. To achieve this goal, our attack needs to fulfill the following requirements:

\begin{itemize}[nosep]
    \item \textbf{Stealth:} A key requirement of our attack is the stealthiness against human vision. Therefore, we use \acp{LED} in the near-infrared spectrum. In contrast to existing attacks in this spectrum~\cite{sato_invisible_2024, wang_invisible_2024}, we do not operate with laser diodes that can harm the human eye, but with low-power \acp{LED} that are invisible to human perception.
    \item \textbf{Dynamic:} To be effective against different target assets and different \ac{ML} models, our \povd{} must be able to show dynamic adaptable content.
    \item \textbf{Triggered:} In contrast to printed physical adversarial patches~\cite{eykholt_robust_2018}, our attack can be triggered to be effective against only specific vehicles. By only switching off the \acp{LED} of our \povd{}, but not the rotation, the traffic sign is still visible, with only negligible occlusion.
    \item \textbf{Remote:} Our attack does not require physical access to the target vehicle, but the \povd{} is attached to an asset, e.g., a traffic sign, and can be controled remotely or without manual intervention.
    \item \textbf{Optimized Position:} While the displayed content and position of the \povd{} can be freely chosen, we optimize our attack for the optimal position that shows the highest effect in our digital simulation.
\end{itemize}

Our attack targets illumination conditions, under which \autoref{eq:povConstraint} is fulfilled. \textbf{This will result in a limited application use case of dawn or nighttime attacks for typical cameras and a reasonable rotation speed of the \povd{}.} Additionally, this limitation allows the use of low-power infrared \acp{LED} that are commercially available, instead of expensive and potentially dangerous laser diodes.

\subsection{Attacker Knowledge and Capabilities}

We assume an attacker \textbf{without prior knowledge of the \ac{ML} model internal weights} ("black-box attack") but with general awareness of the system. Specifically, the attacker needs to know whether the targeted vehicle uses cameras without an infrared cutoff filter. Such generic system knowledge can be obtained from public sources or reference cameras and is similar to assumptions of other physical adversarial attacks~\cite{zhu_tpatch_2023, xia_moire_2024, sato_invisible_2024, man_remote_2024}.

As specified in our requirements, \textbf{the attacker does not need to be physically present during the attack execution} but only for the attack preparation, namely for mounting the \povd{} at the targeted traffic sign. This allows the attack to not raise any suspicion by avoiding the permanent physical presence of the attacker.

Lastly, we assume \textbf{an attacker who has basic knowledge of electrical engineering and access to low-cost commodity hardware} (e.g., infrared LEDs, a microcontroller, and a power supply). While \povd{}s in the human-visible light spectrum are commercially available, infrared ones have to be crafted individually, but can be done with limited knowledge and resources.

\section{Attack Design}
\label{sec:attackDesign}

\begin{figure*}[t]
    \centering
    \includegraphics[width=1\linewidth]{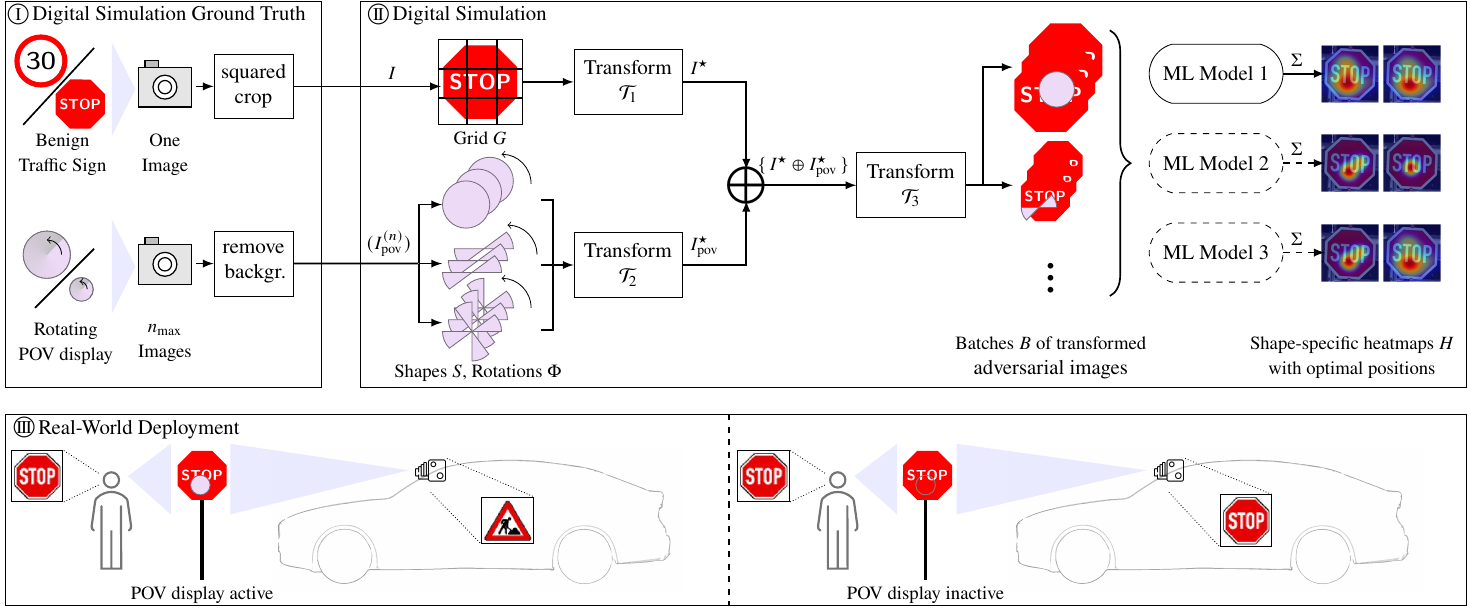}
    \caption{Overview of our \povd{}-based attack. The complete process consists of three steps: The data captured in the \textbf{\circledRomanText{1}~digital simulation ground truth} serves as input to identify the optimal placement of the \povd{} in a \textbf{\circledRomanText{2}~digital simulation}. By \textbf{\circledRomanText{3}~deploying the identified position in the real-world}, and activating the infrared \povd{}, \ac{ML}-based image classification models can be tricked while being invisible to humans.}
    \label{fig:attack_overview}
\end{figure*}

In this section, we will introduce the requirements for our initial prototype of a near-infrared \povd{} and present in detail the steps of our attack based on three stages that are depicted in \autoref{fig:attack_overview}. Our physical attack is based on a digital simulation (\circledRomanText{2}) that requires some initially captured real-world ground-truth images (\circledRomanText{1}) and results in an optimal placement position for the deployment in the real world (\circledRomanText{3}).

\subsection{Infrared POV Display Requirements}

We will present our two-blade near-infrared \povd{} to capture ground-truth images that serve as a basis for our digital simulation. While \povd{}s in the human-visible light spectrum are used for entertainment or advertising purposes and are readily available, to the best of our knowledge, near-infrared \povd{}s are not commercially available. Since the near-infrared spectrum is an essential factor in creating a stealth attack, we craft a prototypical \povd{} hardware. Further details on the configuration and software settings will be explained later in Section~\ref{sec:evaluation}.

The most important aspect in the design of a near-infrared \povd{} is the selection of infrared \acp{LED}. The sensitivity of an image sensor to a specific wavelength is defined by the quantum efficiency~\cite{el_gamal_cmos_2005, yadid-pecht_fundamentals_2004}. While the exact quantum efficiency curve depends on the used image sensor in the camera, \ac{CMOS}-based color image sensors typically show a local maximum between approx. 800nm to 900nm~\cite{perkins_near_2023, hamamatsu_photonics_kk_ccdcmos_2025, allied_vision_technologies_gmbh_alvium_2025}. This requirement limits the search for suitable \acp{LED} to this wavelength spectrum. In our prototypical implementation, we, therefore, use near-infrared \acp{LED} with a wavelength of 860nm~\cite{kingbright_wp7113sf6bt-p22_2024}. These \acp{LED} can be mounted on a \ac{PCB}, attached to a small electric motor. Different patterns can then be visualized by using \ac{PWM} to create rotation-symmetric circular sectors.

\subsection{\circledRomanText{1} Digital Simulation Ground Truth}

Our digital simulation of the \ac{POV} attack requires two inputs:

\begin{enumerate}
    \item A single benign image of a traffic sign $I$ which shows the traffic sign that shall be attacked.
    \item A sequence of $n_{\text{max}}$ images of a rotating infrared \povd{} with a given diameter, displaying a full circle $(I_{\mathrm{pov}}^{(n)})_{n=1}^{n_{\text{max}}}$, acting as a basis for the digitally simulated \povd. For our prototypical implementation, we will select $n_{\text{max}}=10$.
\end{enumerate}

Both inputs are optimally captured on a neutral background, such as a black wall. Afterwards, we crop the image of the traffic sign to result in a square aspect ratio required by the targeted image classification model. From the $n_{\text{max}}$ images of the rotating infrared \povd, we remove the background by setting it transparent, showing only the full circle. While our digital simulation can also work with only a single image of a \povd, multiple consecutive images compensate for flickering artifacts and changing light intensities due to potentially overlapping regions in the \povd, as shown in \autoref{fig:pov_effect_texp_greater_frot}. This makes the digital simulation more robust.

\subsection{\circledRomanText{2} Digital Simulation}

The goal of the digital simulation is the identification of the optimal position of the \povd{} on the targeted traffic sign to ensure high attack success for the \ac{ML} model under test. We use the prepared ground truth captured in \circledRomanText{1} to create model-specific heatmaps that visualize the most successful attack positions. A detailed description is available in Algorithm~\ref{img:digitalSimulationAlgorithm}.

In a first step, we apply a grid $G=\{0{:}m\}\times\{0{:}m\}$ on the traffic sign where each grid point marks a possible placement option of the \povd. Additionally, we create different shapes $S$, representing rotation-symmetric circular sectors that the \povd{} can display with low-level control hardware that can easily be deployed in the real world. Different numbers of sectors can be visualized with a two-blade \povd{}. Additionally, we create a configurable number of rotated circular sectors $\Phi$ to compensate for the dynamic changes of the rotating \povd. While the ground-truth images from \circledRomanText{1} are required for each new attack, the grid, shape definition, and rotated circular sectors are configurable parameters that can be fixed once by the attacker.

\begin{algorithm}[t]
\caption{Digital simulation from benign traffic-sign image $I$ and a $n_{\text{max}}$-frame \povd{} sequence $I_{\mathrm{pov}}^{(1{:}n_{\text{max}})}$.}
\label{img:digitalSimulationAlgorithm}
\includegraphics[width=\linewidth]{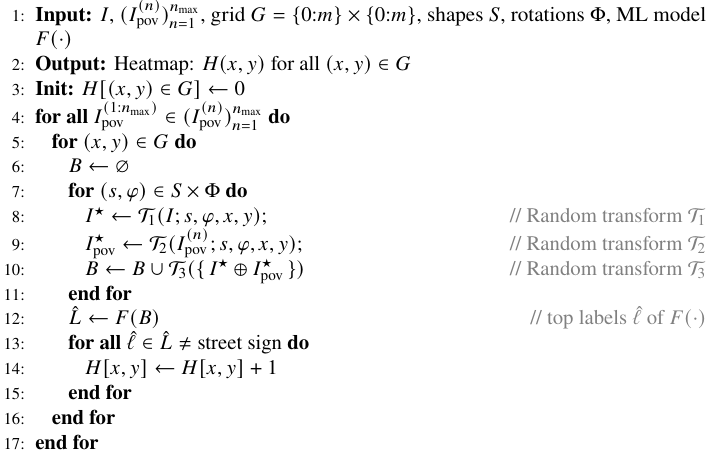}
\end{algorithm}

\begin{figure}[t]
     \centering
     \begin{subfigure}[b]{0.32\linewidth}
         \centering
         \includegraphics[width=0.9\textwidth]{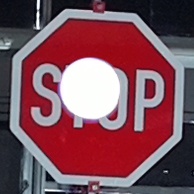}
         \caption{Full \povd{} without transformations}
         \label{fig:dt_example_original_fullcircle}
     \end{subfigure}
     \hfill
     \begin{subfigure}[b]{0.32\linewidth}
         \centering
         \includegraphics[width=0.9\textwidth]{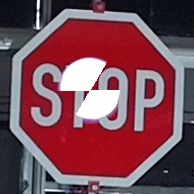}
         \caption{Two circular sectors without transformations}
         \label{fig:dt_example_original_2slices}
     \end{subfigure}
     \hfill
     \begin{subfigure}[b]{0.32\linewidth}
         \centering
         \includegraphics[width=0.9\textwidth]{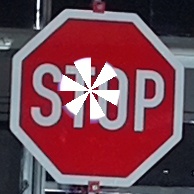}
         \caption{Six circular sectors without transformations}
         \label{fig:dt_example_original_6slices}
     \end{subfigure}
     \hfill
     \begin{subfigure}[b]{0.32\linewidth}
         \centering
         \includegraphics[width=0.9\textwidth]{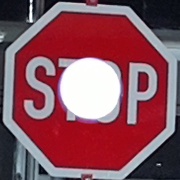}
         \caption{Full \povd{} with crop transformation}
         \label{fig:dt_example_transformCrop_fullcircle}
     \end{subfigure}
     \hfill
     \begin{subfigure}[b]{0.32\linewidth}
         \centering
         \includegraphics[width=0.9\textwidth]{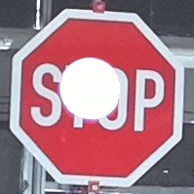}
         \caption{Full \povd{} with brightness transformation}
         \label{fig:dt_example_transformBrighter_fullcircle}
     \end{subfigure}
     \hfill
     \begin{subfigure}[b]{0.32\linewidth}
         \centering
         \includegraphics[width=0.9\textwidth]{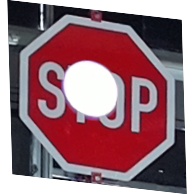}
         \caption{Full \povd{} with two perspective transformations}
         \label{fig:dt_example_transformTilt_fullcircle}
     \end{subfigure}
     \caption{Example images from the digital simulation. All images show the overlay of the captured \povd{} images and the benign traffic sign with different transformations.}
     \label{fig:dt_example}
\end{figure}

For both the benign traffic sign and the different shapes of the \povd, we apply transformations $\mathcal{T}_1,\mathcal{T}_2$ individually (lines 8 and 9), similar to existing approaches of adversarial attacks~\cite{athalye_synthesizing_2018}. After overlaying the different shapes on the grid positions of the benign traffic sign, we apply additional transformations $\mathcal{T}_3$ to the combined resulting images $I^\star \oplus I_{\mathrm{pov}}^\star$ (line 10). \autoref{tab:digitalSimulationTransformation} shows the various available transformation parameters, including their possible values, and where they are applied. The exact values are randomly sampled from the specified range. By applying these randomly selected transformations with randomly sampled values, the resulting placement options in the digital simulation are more robust to environmental factors such as different illumination or changing perspectives that occur in the real world. Additionally, \povd{}-specific parameters, such as the scaling, can help to identify the necessary size of \povd{}s for real-world deployment. A visual depiction of some representative transformations is available in \autoref{fig:dt_example}.

\begin{table}[t]
\centering
\caption{List of available transformations, including their value range and the stage they are applied. Brightness transformations are based on the YCbCr color model.}
\label{tab:digitalSimulationTransformation}
\resizebox{1\linewidth}{!}{%
    \begin{tabular}{cccc}\toprule
        Transformation & Stage & Parameter & Possible Range \\ \toprule
        \multirow{2}{*}{Brightness} & Traffic sign & $\Delta\text{Y}$ in DN & $-50\dots 50$ \\
         & \povd & $\Delta\text{Y}$ in DN & $20\dots 50$ \\
        \multirow{2}{*}{Scale} & \povd & Factor & $0.3\dots 1.5$ \\
         & Combined image & Factor & $0.5\dots 1.2$ \\
        \multirow{3}{*}{Perspective} & \multirow{3}{*}{Combined image} & Tilt\textsubscript{horizontal} in $^\circ$ & $5\dots 30$ \\
        & & Tilt\textsubscript{vertical} in $^\circ$ & $5\dots 30$ \\
        & & Perspective Factor & $0.05\dots 0.10$ \\
        \multirow{2}{*}{Glare} & \multirow{2}{*}{\povd} & Extension Factor & $1.2\dots 1.5$ \\
        & & Fade Strength & $0.2\dots 0.4$ \\
        Center Crop & Combined image & Crop Factor & $0.01\dots 0.05$ \\\toprule
\end{tabular}}
\end{table}

The resulting batches $B$ of transformed, digitally simulated, adversarial images serve as an input to one or multiple targeted \ac{ML} image classifiers $F(\cdot)$ (line 12). Since we do not apply backpropagation and do not work with the model's internal weights or architecture, our attack is a black-box attack. For each digitally simulated adversarial image, we run the \ac{ML} model inference and store only the top label $\hat{\ell}$, the top confidence, the confidence of the class "street sign", and the respective configuration of our digital simulation, including the applied transformation parameters and shape configuration.

We aggregate the number of misclassified frames for each grid position based on sequential frames, applied rotations of the circular sectors, and applied transformations, resulting in a two-dimensional heatmap $H$ that can be overlayed on the original image to identify the optimal placement position of the \povd{} (line 14).

\subsection{\circledRomanText{3} Real-World Deployment}

The final stage of the attack is the deployment of the infrared \povd{} in the real world by placing it at the position of the resulting heatmap $H$ from the digital simulation. By using \ac{PWM}, circular sectors as shown in Figures~\ref{fig:dt_example_original_2slices} and~\ref{fig:dt_example_original_6slices} can be achieved even with low-cost commodity \acp{LED}. With the transformations applied in the digital simulation, the resulting optimal position already compensates possible deviations of environmental conditions during deployment.

\section{Evaluation}
\label{sec:evaluation}

We evaluate both the digitally simulated images and the real-world deployment of the \povd. In our initial overview, we will define our deployment setup and the attack success metrics.

\subsection{Overview}

\subsubsection{Hardware Setup}
\begin{figure}[t]
     \centering
     \begin{subfigure}[b]{0.32\linewidth}
         \centering
         \includegraphics[width=1\linewidth]{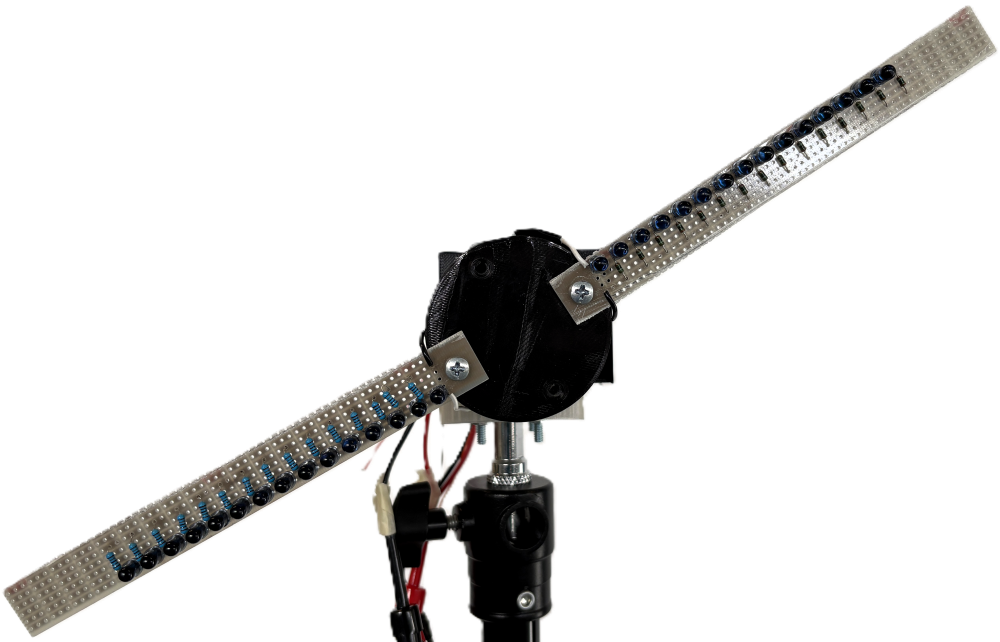}
         \caption{\povd{} with 30cm diameter mounted on a tripod}
         \label{fig:povd_prototype_large}
     \end{subfigure}
     \hfill
     \begin{subfigure}[b]{0.32\linewidth}
         \centering
         \includegraphics[width=1\linewidth]{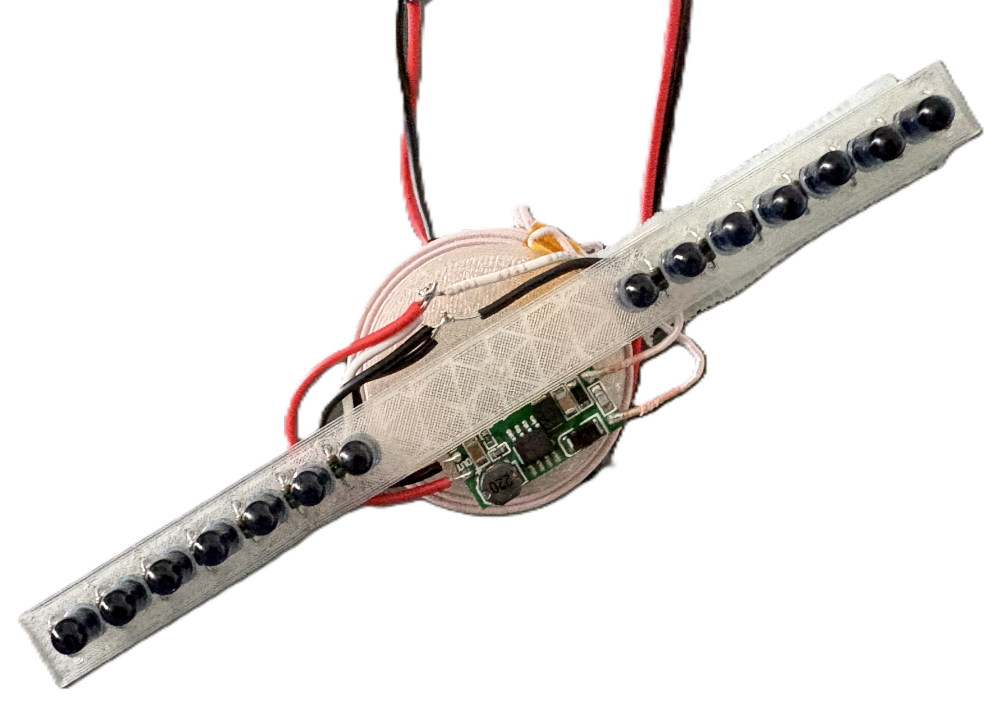}
         \caption{\povd{} with 15cm diameter for mobile deployment}
         \label{fig:povd_prototype_portable}
     \end{subfigure}
     \hfill
     \begin{subfigure}[b]{0.32\linewidth}
         \centering
         \includegraphics[width=0.8\linewidth]{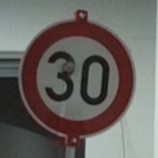}
         \caption{Mobile \povd{} with 15cm diameter on a traffic sign}
         \label{fig:hologram_prototype_portable_ledOff}
     \end{subfigure}
    \caption{Prototypical hardware setups of a near-infrared \povd. The \acp{LED} are soldered with their respective resistors on a \ac{PCB}, which is attached to the motor. The portable \povd{} allows for stealth deployment, if the \acp{LED} are switched off and the fan is rotating.}
    \label{fig:povd_prototype}
\end{figure}

In contrast to existing infrared attacks~\cite{sato_invisible_2024, wang_i_2021}, we do not work with highly directed infrared lasers but commercially available near-infrared \acp{LED} with a wavelength of 860nm~\cite{kingbright_wp7113sf6bt-p22_2024}, thus making it stealth and less harmful for the human eye. All \acp{LED} are soldered in parallel on a \ac{PCB} to limit the required voltage for operation. The total diameter of our \povd{} is 30cm, but larger and smaller hardware is possible. For selected evaluations, we also test \povd{}s with diameters of 10cm and 20cm. We use a direct current motor~\cite{handson_technology_775_2025}, operated at 12V, to rotate the \acp{PCB} with the \acp{LED}. While different motors can be used, it is important to note that the timing constraints of \autoref{eq:povConstraint} must be fulfilled, even with the \acp{PCB} and \acp{LED} attached. For the power transmission of the stationary power supply to the rotating \acp{LED}, we use slip rings. Figure~\ref{fig:povd_prototype_large} shows the developed prototype of our near-infrared \povd{} with a diameter of 30cm that we also use in our evaluation.

Additionally, we show a 15cm version of the near-infrared \povd{} that allows for portable deployment. It can be directly attached to traffic signs without the need for an external mounting. We use a 3D-printed body that contains the motor~\cite{motraxx_elektrogerate_gmbh_fk-280sav-19170_2022} and blades with the near-infrared \acp{LED}~\cite{kingbright_wp7113sf6bt-p22_2024}. This body is connected through neodymium magnets on both sides of the traffic sign, since the sign itself is not magnetic. The wires for the power supply can be reduced to a minimum by only leading to the back of the traffic sign, where a battery can be hidden. With this setup, the attacker can deploy and remove the \povd{} very fast to ensure high stealthiness. Figures~\ref{fig:povd_prototype_portable} and \ref{fig:hologram_prototype_portable_ledOff} show the portable version of our \povd{} and highlight its stealthiness if the \acp{LED} are off.

As a camera, we select an embedded camera module featuring a Sony IMX708 image sensor with no infrared filter~\cite{raspberry_pi_ltd_camera_2024}. This camera module has a wide field of view and comparable features, including high dynamic range imaging, to those of real automotive image sensors.

\subsubsection{Targeted Traffic Signs}
Since our attack operates at the sensor level by displaying controlled infrared patterns received by the camera, it is sign-agnostic and does not rely on the semantic properties of individual traffic signs. For evaluation, we focus on the following traffic signs:

\begin{enumerate}[nosep]
    \item A stop sign, as this sign can be classified/detected by all evaluated \ac{ML} models.
    \item A German 30km/h speed limit sign, as \ac{GTSRB}-trained \ac{ML} models can classify this sign specifically. Additionally, it represents a differently sized and shaped traffic sign.
\end{enumerate}

Both signs are utilized in widely used benchmarks and prior physical adversarial attack studies~\cite{stallkamp_man_2012, eykholt_robust_2018} and represent safety-critical traffic signs with distinct visual characteristics, enabling controlled and reproducible evaluation.

\subsubsection{Perception-Level Evaluation}
We aim to evaluate the impact of our \povd{} attack at the perception level and, therefore, do not perform an end-to-end validation on a specific commercial vehicle. End-to-end autonomy stacks differ substantially across platforms in their planning and control logic, making vehicle-level behavior highly implementation-dependent~\cite{wang_revisiting_2025}. By focusing on perception models, we isolate the effect of the attack on a core component shared across autonomous driving systems while still mimicking an approaching vehicle through physical evaluations at distances up to 20m. Prior work has shown that perception-level attacks can propagate to system-level safety risks~\cite{wang_revisiting_2025}.

We will evaluate all steps of the attack design from Section~\ref{sec:attackDesign} independently for two different \ac{ML}-based image classification models
\begin{enumerate*}[before=\unskip{: }, itemjoin={{; }}, itemjoin*={{, and }}, label={(\roman*)}]
\item \textit{ResNet-50}~\cite{he_deep_2016}
\item \textit{ConvNeXt small}~\cite{liu_convnet_2022}. 
\end{enumerate*}
Both models were trained on the \ac{GTSRB} dataset~\cite{stallkamp_man_2012}, using the same training parameters as the pretrained models deployed by PyTorch\footnote{\url{https://github.com/pytorch/vision/tree/main/references/classification}}. Additionally, we analyze the \textbf{transferability} of our attack across \textbf{12 image classification and detection models}, as shown in \autoref{tab:mlModelsTransferability}. We analyze the transferability of the stop sign using both \ac{GTSRB}- and pretrained versions of the \ac{ML} models (ImageNet~\cite{deng_imagenet_2009}, or COCO~\cite{fleet_microsoft_2014}). However, we analyze the transferability of the speed limit sign only against the \ac{GTSRB}-trained versions, since only these models can recognize this specific sign. For all models, we will show benign performance with a switched-off \povd{} because the results are similar to those without any \povd.

\begin{table}[t]
\centering
\caption{List of models and their respective training datasets used for evaluation of the physical deployment of our attack.}
\label{tab:mlModelsTransferability}
\resizebox{0.7\linewidth}{!}{%
    \begin{tabular}{cl}\toprule
        \textbf{Training Dataset} & \textbf{Model} \\ \toprule
        \multirow{7}{*}{\ac{GTSRB}~\cite{stallkamp_man_2012}} & ConvNeXt small~\cite{liu_convnet_2022} \\
         & ConvNeXt base~\cite{liu_convnet_2022} \\
         & ResNet-50~\cite{he_deep_2016} \\
         & ResNet-152~\cite{he_deep_2016} \\
         & ResNext-101~\cite{xie_aggregated_2017} \\
         & VGG16~\cite{simonyan_very_2015} \\
         & ViT-32~\cite{dosovitskiy_image_2021} \\ \hline
        \multirow{3}{*}{ImageNet~\cite{deng_imagenet_2009}} & ConvNeXt base~\cite{liu_convnet_2022} \\
         & ResNet-152~\cite{he_deep_2016} \\
         & ResNet-50~\cite{he_deep_2016} \\ \hline
        \multirow{2}{*}{COCO~\cite{fleet_microsoft_2014}} & YOLO11~\cite{jocher_ultralytics_2024} \\
         & Faster R-CNN~\cite{ren_faster_2015} \\\toprule
\end{tabular}}
\end{table}

\subsubsection{Attack Success}
As defined in our threat model, the goal of our attack is to cause misclassifications in \ac{ML}-based image classification models. At the same time, our \povd{} can exhibit temporal artifacts, such as non-perfectly synchronized \ac{PWM} signals and motor speeds. For this reason, we evaluate the physical deployment of our attack over video durations that are sufficient to observe periodic temporal artifacts. In our experiment, we used 30-second video snippets. 
We extract each image from the video and run it through the respective image classification model, defining the \ac{ASR} as the ratio of misclassified frames within each snippet, consistent with existing approaches of sticker-based adversarial attacks~\cite{eykholt_robust_2018}. We emphasize that this definition of the \ac{ASR} already accounts for real-world distortions that may arise during actual deployment.

\subsection{Digital Test}
We perform digital tests for both the stop sign and the German 30km/h speed limit sign using a 30cm \povd{}.  
Since a 30 cm \povd{} covers a large fraction of the 30km/h speed limit sign, we further evaluate \povd{}s with diameters of 10cm and 20cm to study the feasibility of smaller, even stealthier, and easier-to-deploy devices. For both traffic signs, we run steps \circledRomanText{1} and \circledRomanText{2} from the attack design of \autoref{fig:attack_overview} with two representative shapes, namely a full circle (\raisebox{-0.5ex}{\fullcircle{0.8em}}) and two circular sectors (\raisebox{-0.5ex}{\twoslices{0.5em}}), which can be displayed with the \ac{PWM} controlled \acp{LED}. As defined in our overview, we run the digital simulation for both ResNet-50 \ac{GTSRB} and ConvNeXt small \ac{GTSRB} models. 

\subsubsection{Stop Sign}
\begin{figure}[t]
     \centering
     \begin{subfigure}[b]{0.98\linewidth}
         \centering
         \includegraphics[width=0.24\textwidth]{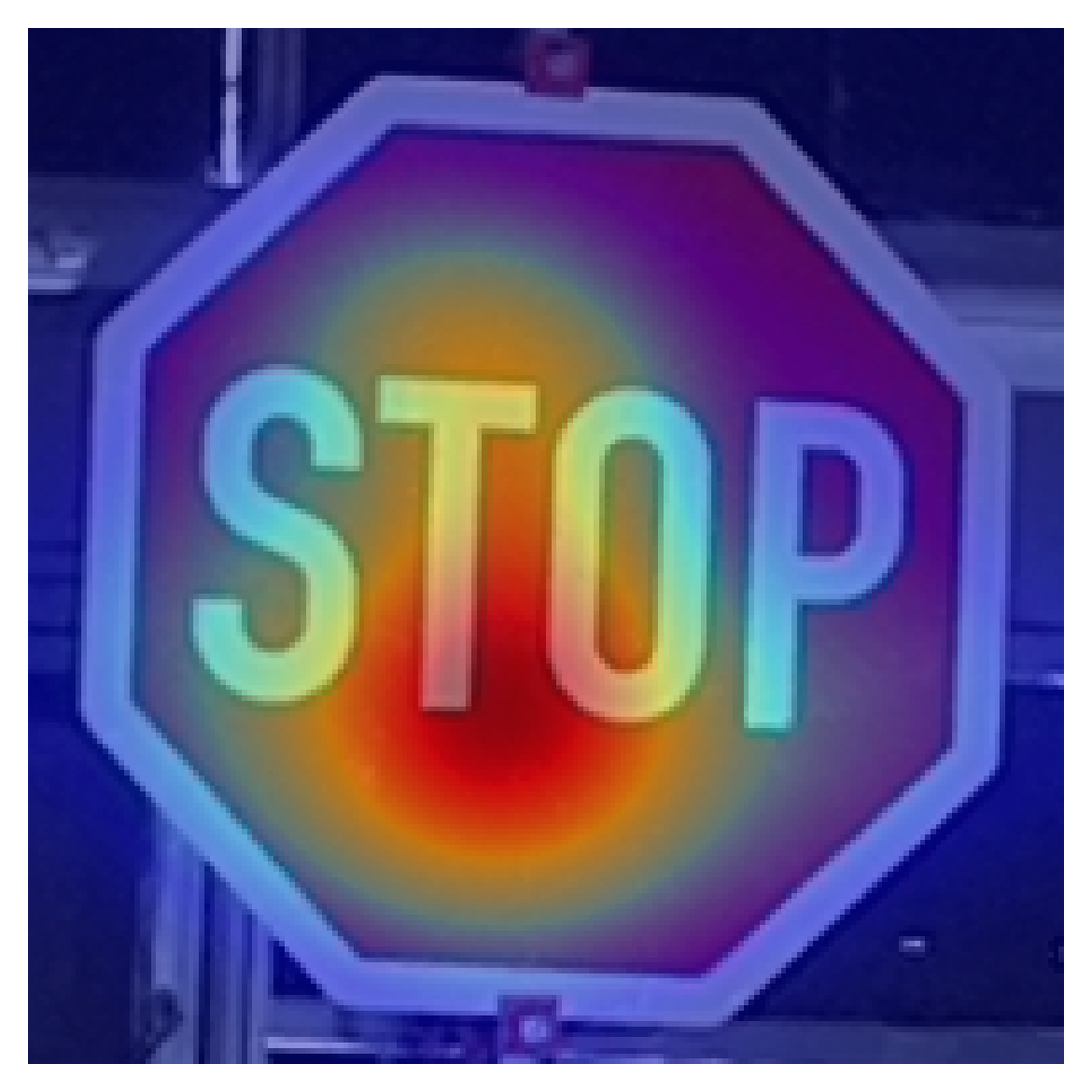}
         \includegraphics[width=0.24\textwidth]{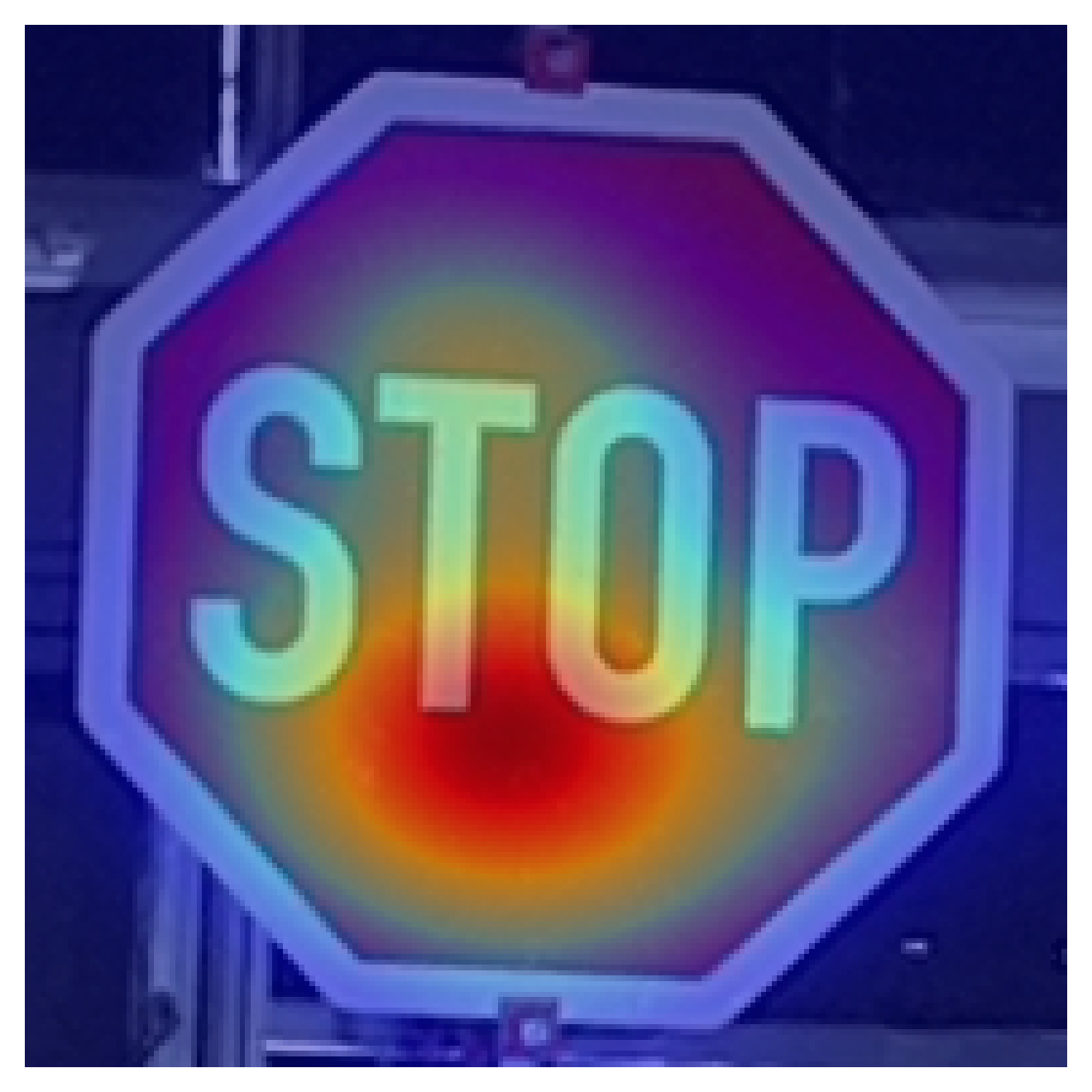}
         \includegraphics[width=0.24\textwidth]{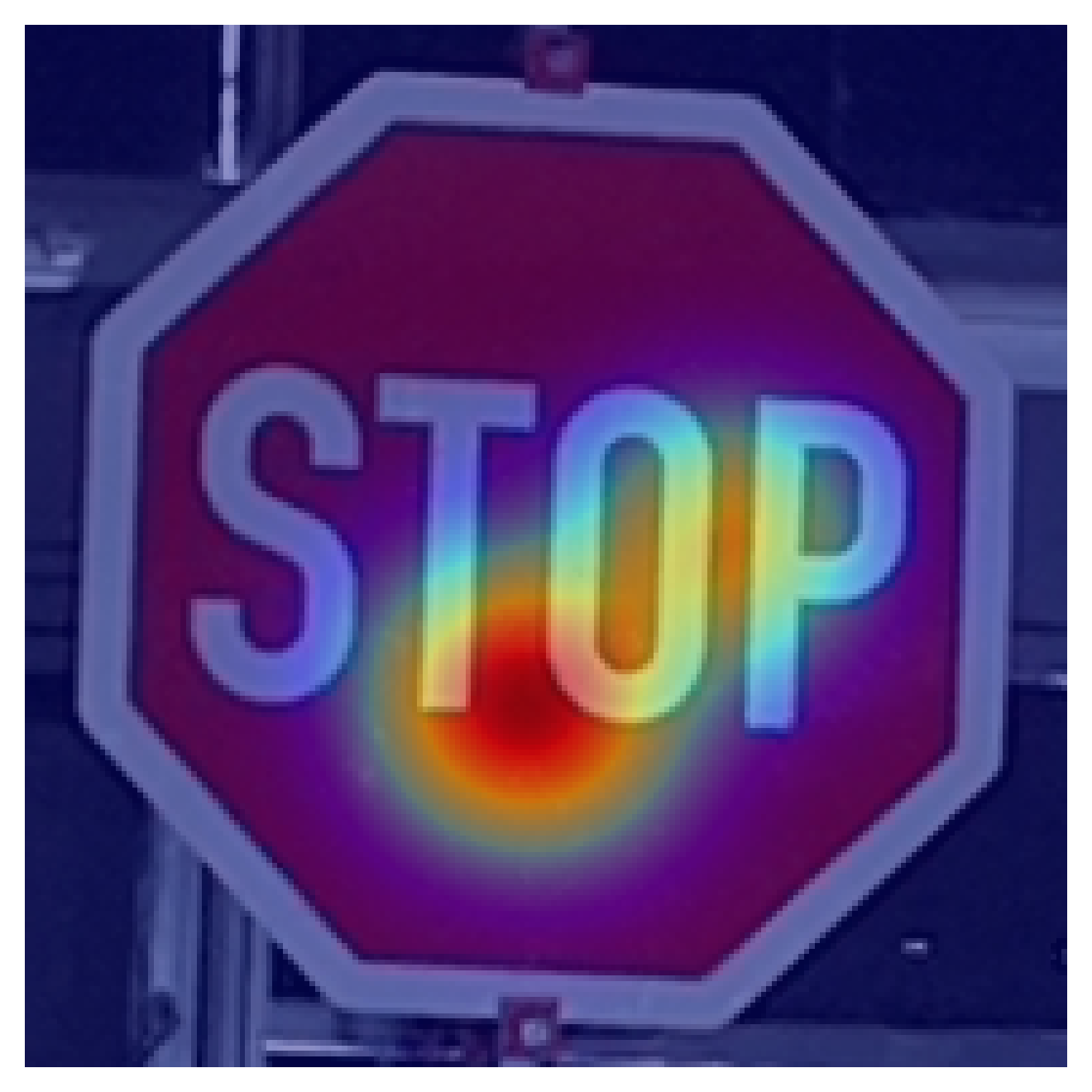}
         \includegraphics[width=0.24\textwidth]{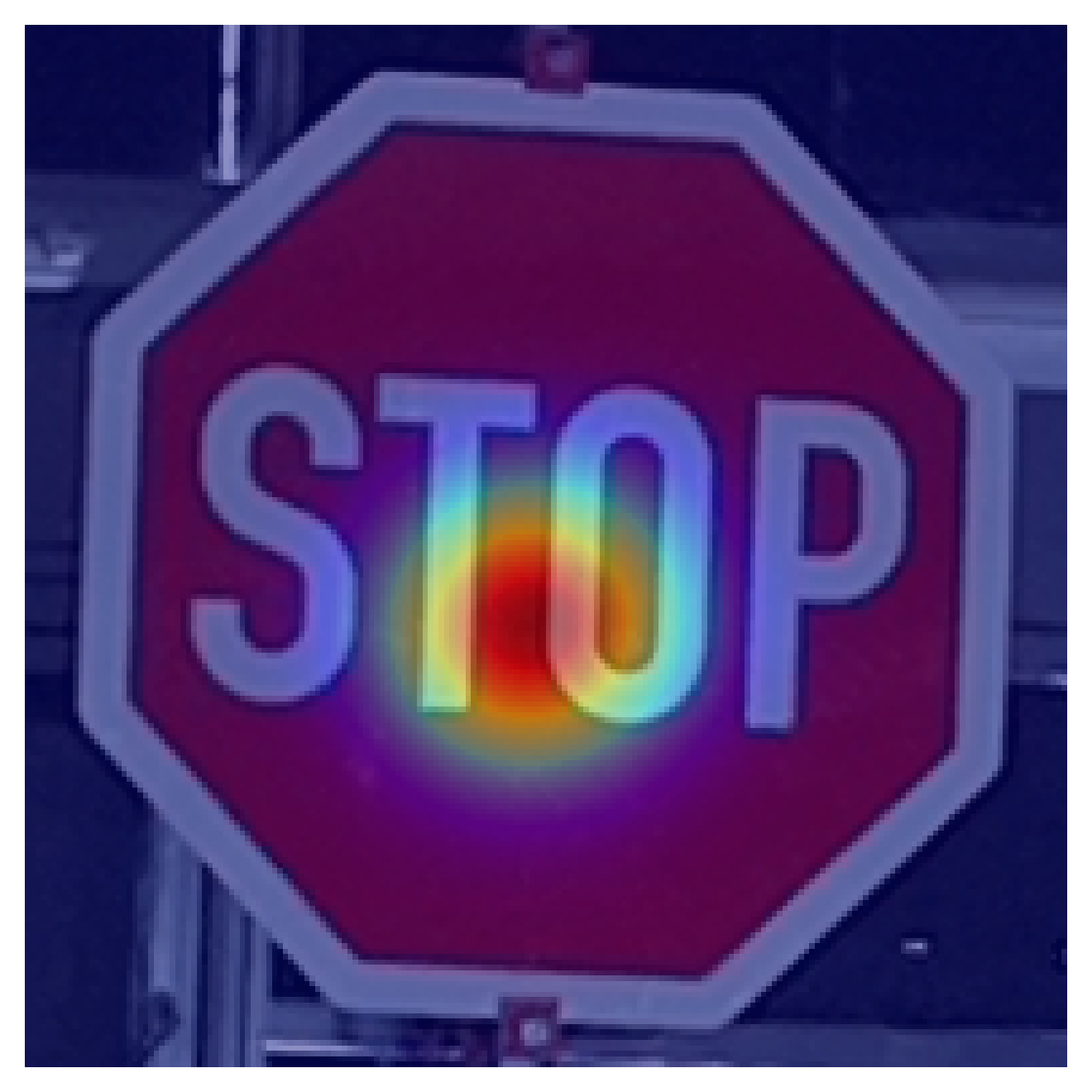}
         \caption{Heatmaps for the stop sign. From left to right\begin{enumerate*}[before=\unskip{: }, itemjoin={{; }}, itemjoin*={{, and }}, label={(\roman*)}]
            \item ResNet-50, full circle
            \item ResNet-50, two sectors
            \item ConvNeXt small, full circle
            \item ConvNeXt small, two sectors.
         \end{enumerate*}}
         \label{fig:heatmaps_stop}
     \end{subfigure}
     \hfill
     \begin{subfigure}[b]{0.98\linewidth}
         \centering
         \includegraphics[width=0.24\textwidth]{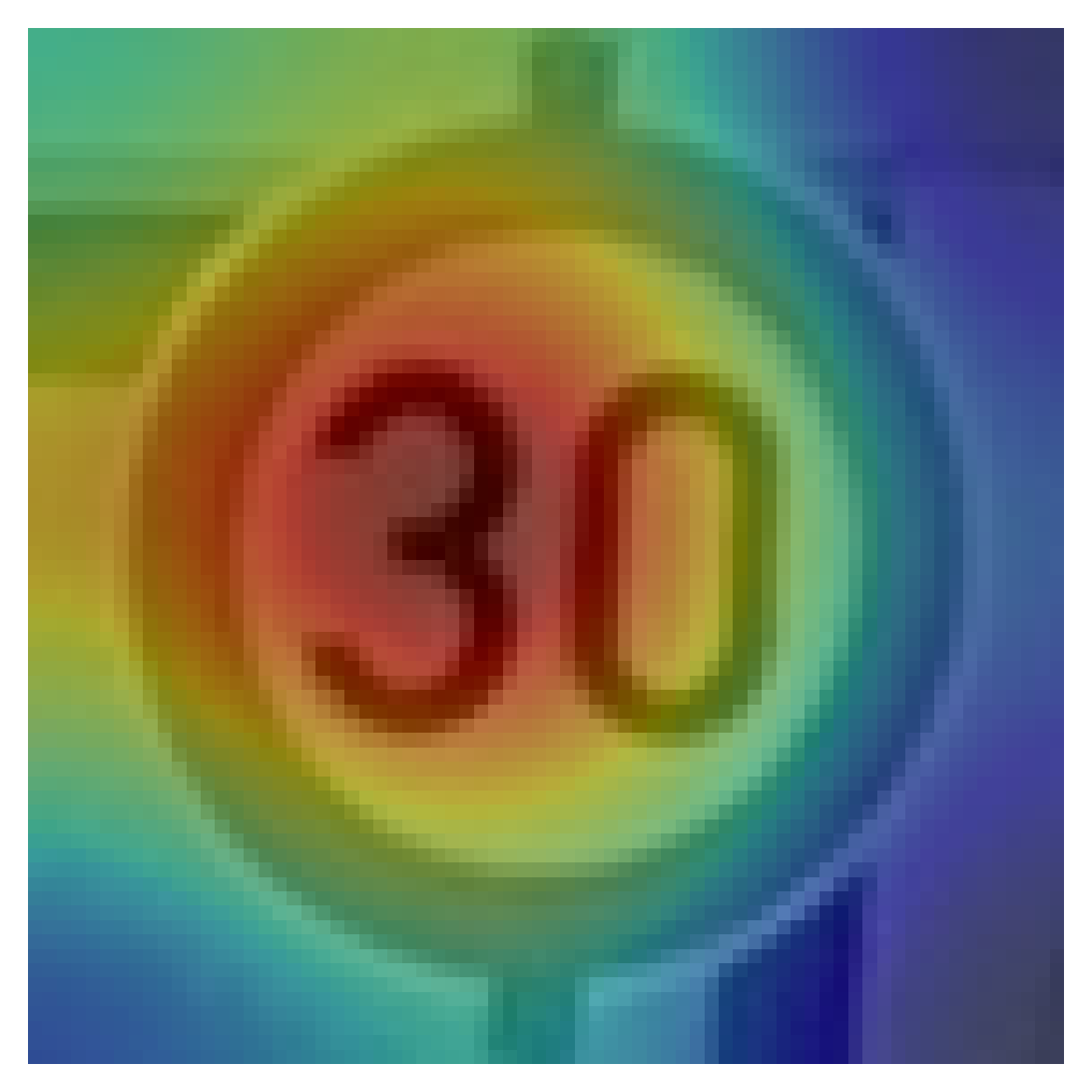}
         \includegraphics[width=0.24\textwidth]{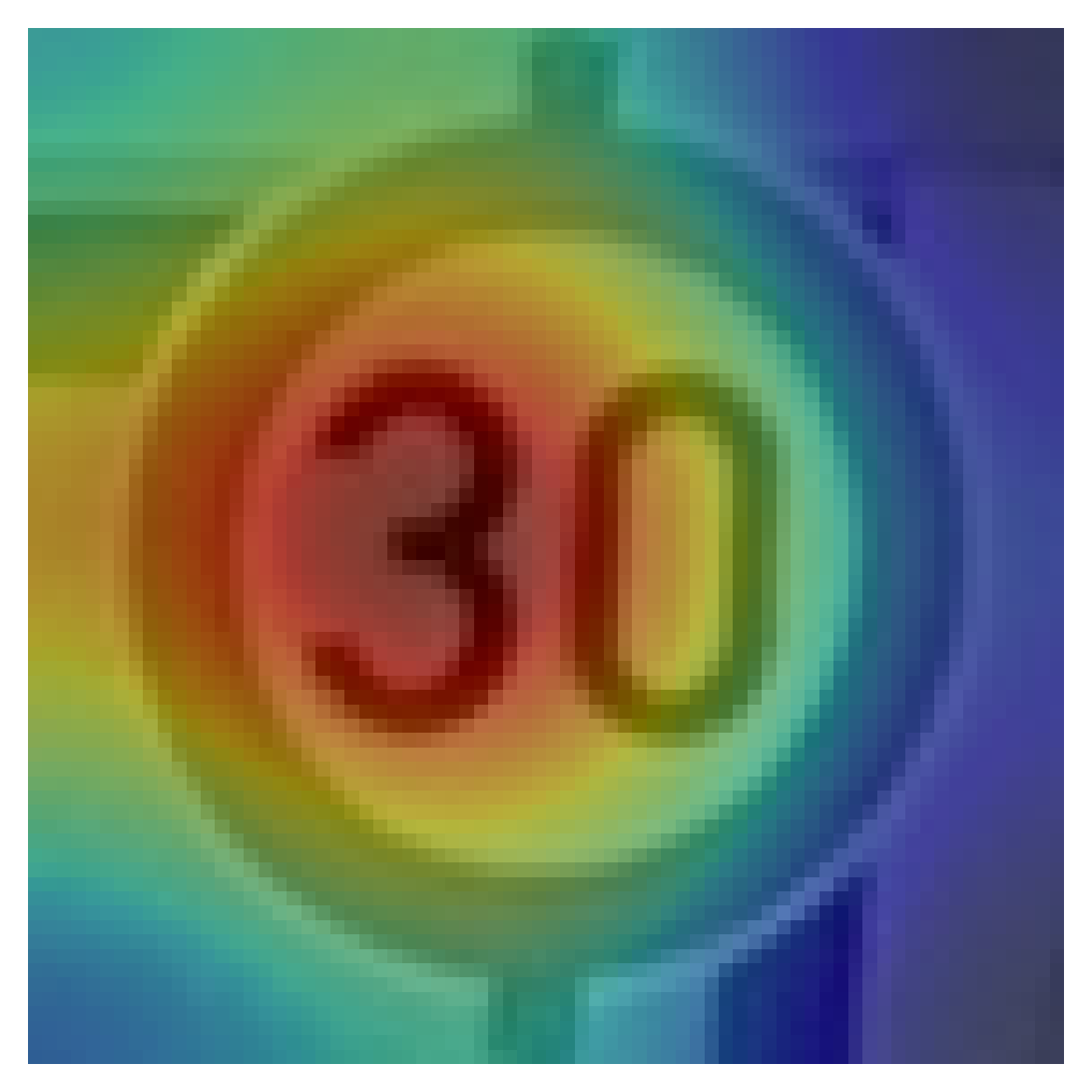}
         \includegraphics[width=0.24\textwidth]{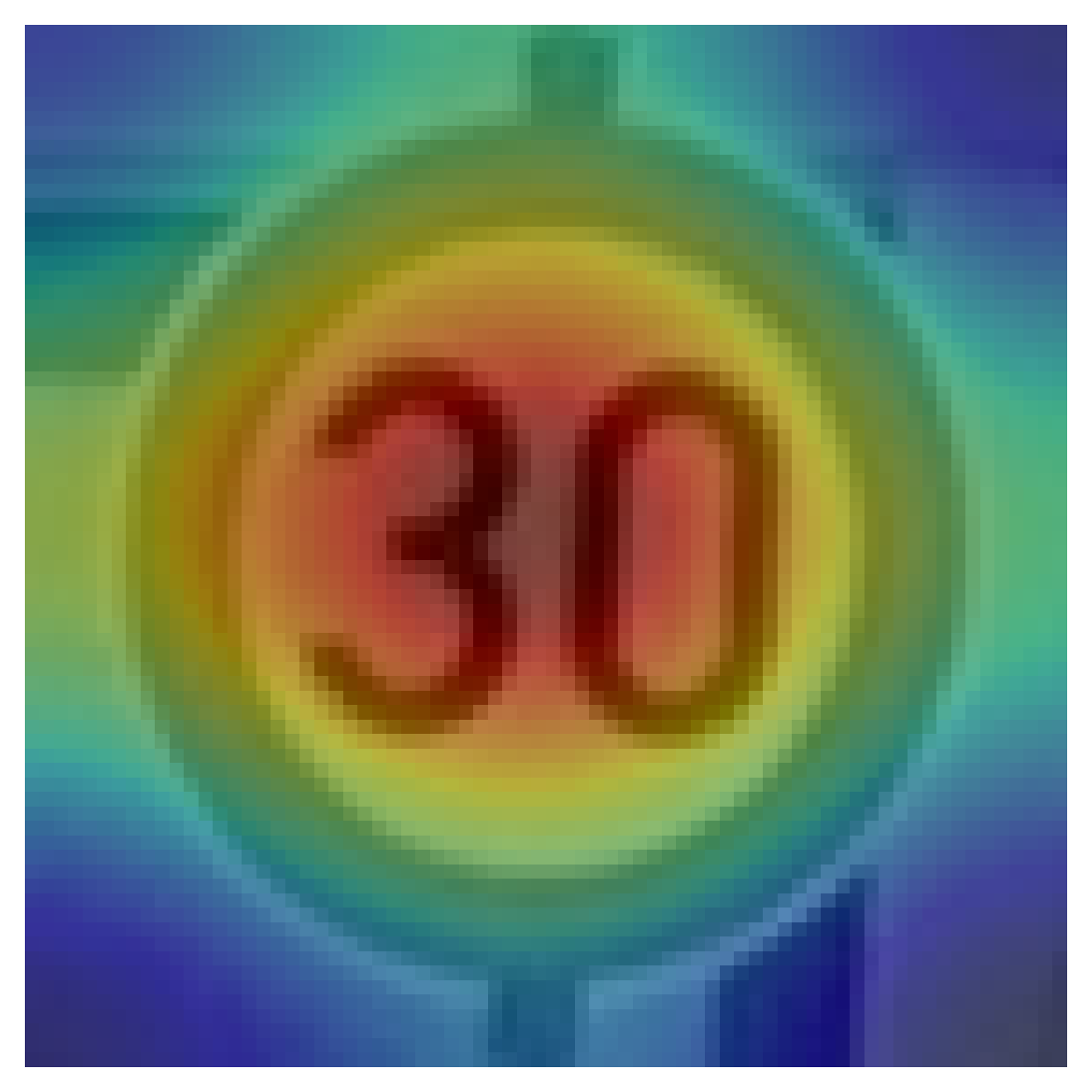}
         \includegraphics[width=0.24\textwidth]{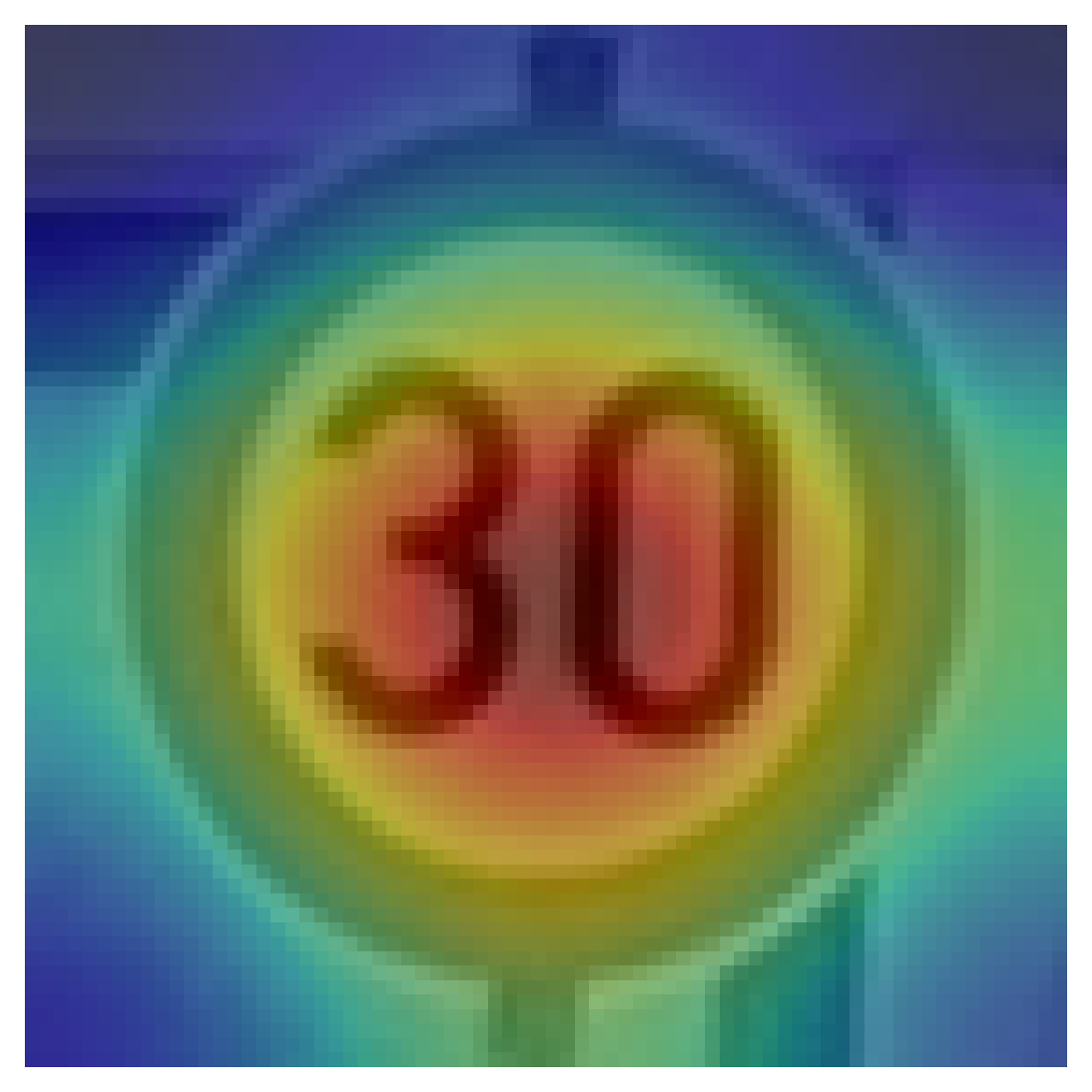}
         \caption{Heatmaps for the 30km/h speed limit sign. From left to right\begin{enumerate*}[before=\unskip{: }, itemjoin={{; }}, itemjoin*={{, and }}, label={(\roman*)}]
            \item ResNet-50, full circle
            \item ResNet-50, two sectors
            \item ConvNeXt small, full circle
            \item ConvNeXt small, two sectors.
         \end{enumerate*}}
         \label{fig:heatmaps_speedlimit}
     \end{subfigure}
     \hfill
     \begin{subfigure}[b]{0.98\linewidth}
         \centering
         \includegraphics[width=0.24\textwidth]{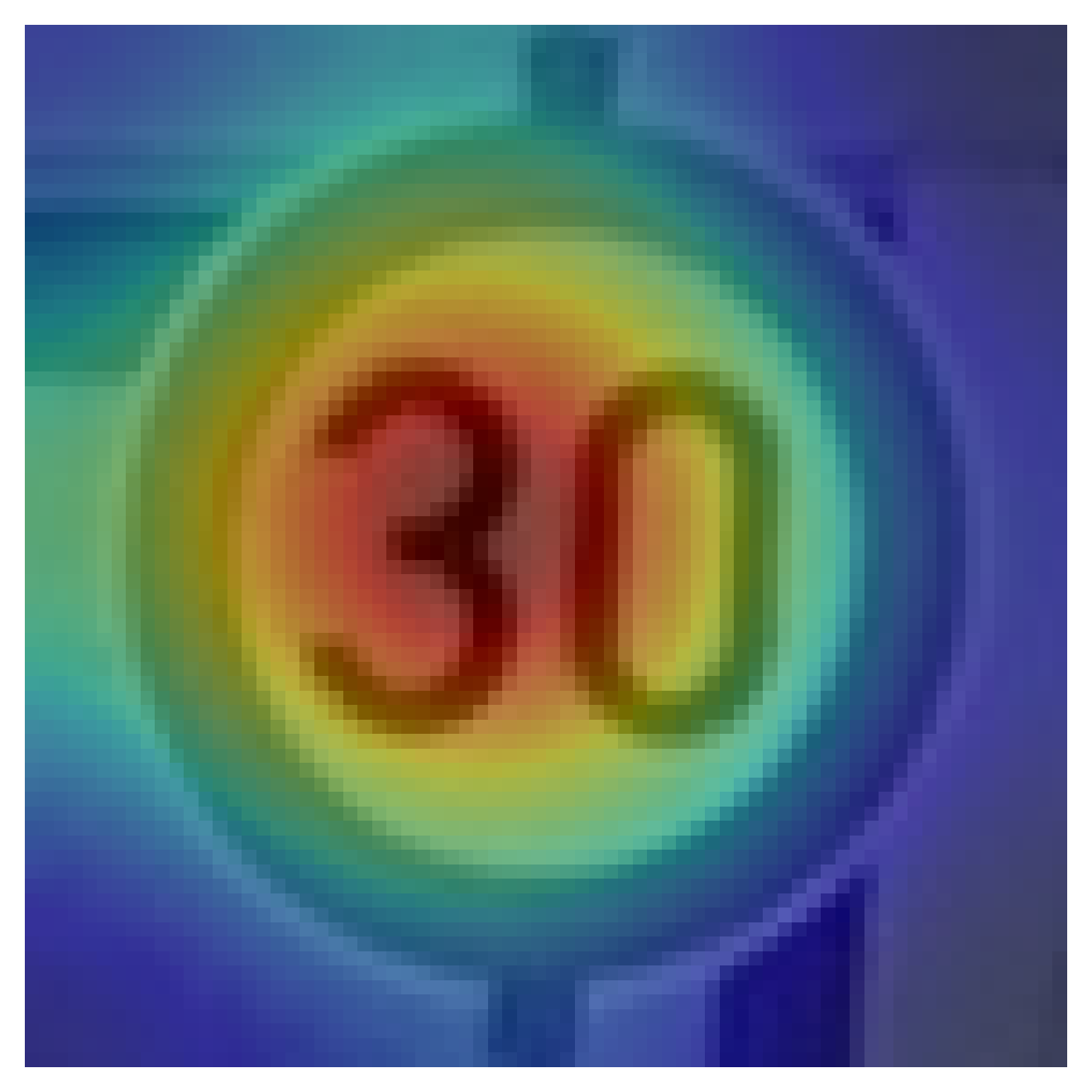}
         \includegraphics[width=0.24\textwidth]{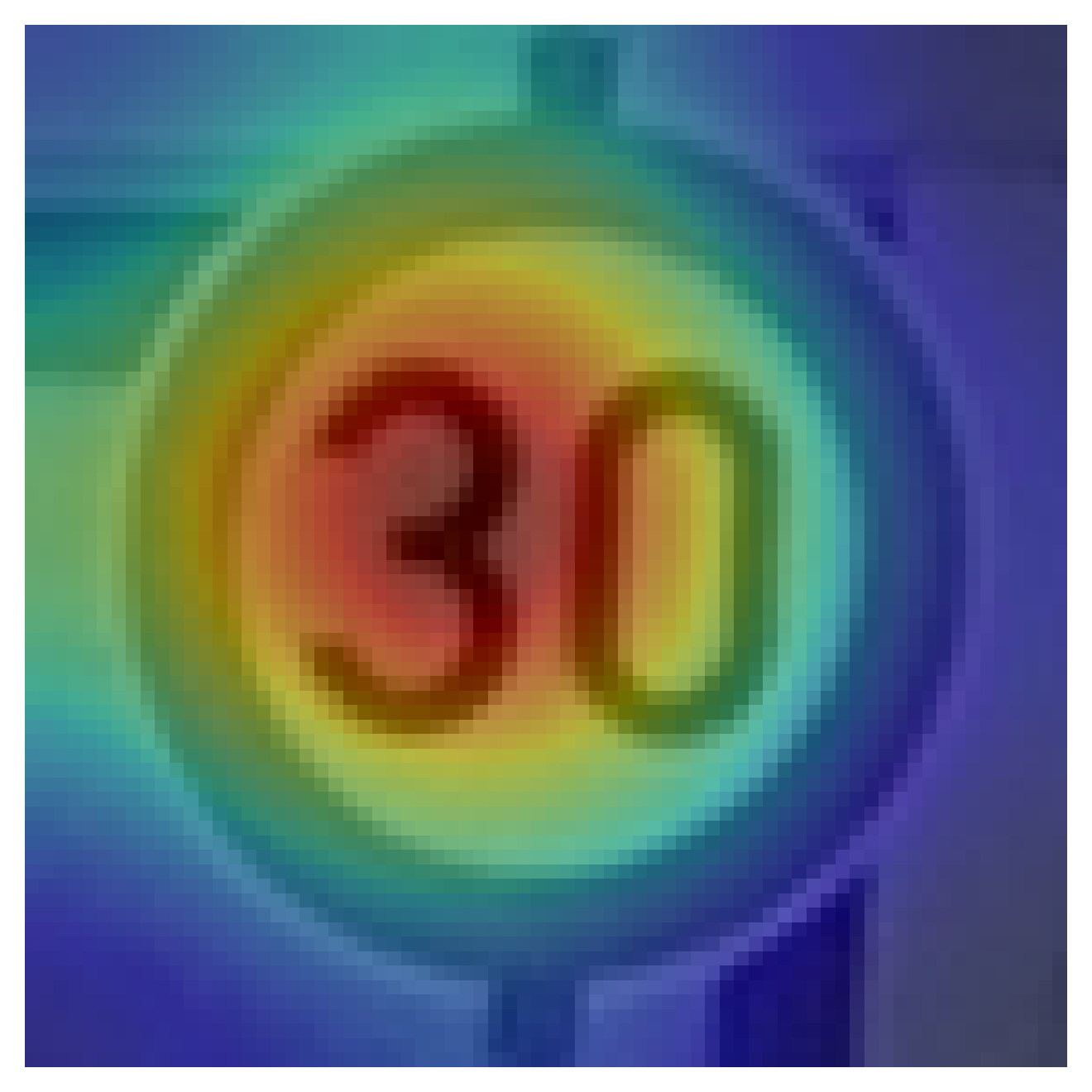}
         \includegraphics[width=0.24\textwidth]{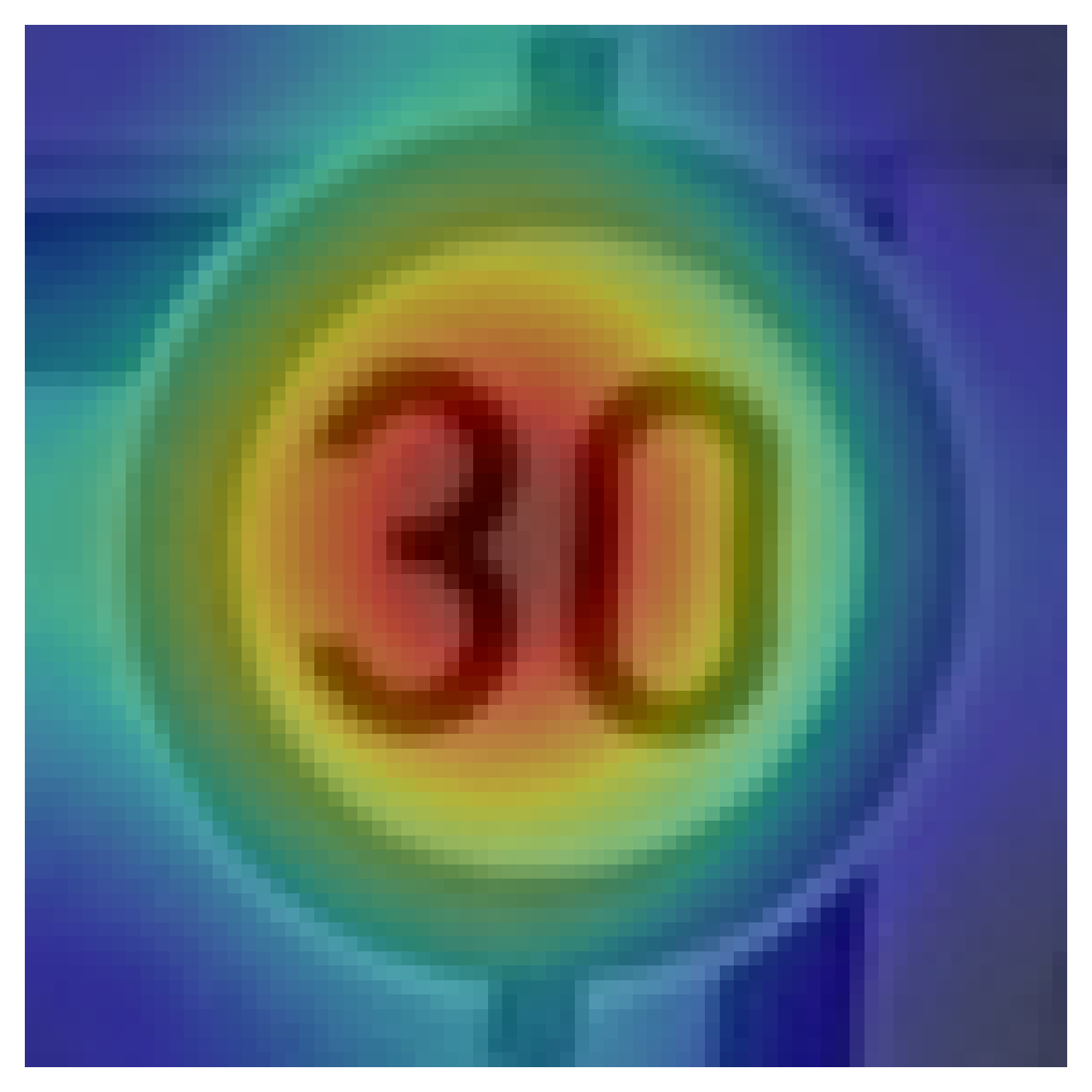}
         \includegraphics[width=0.24\textwidth]{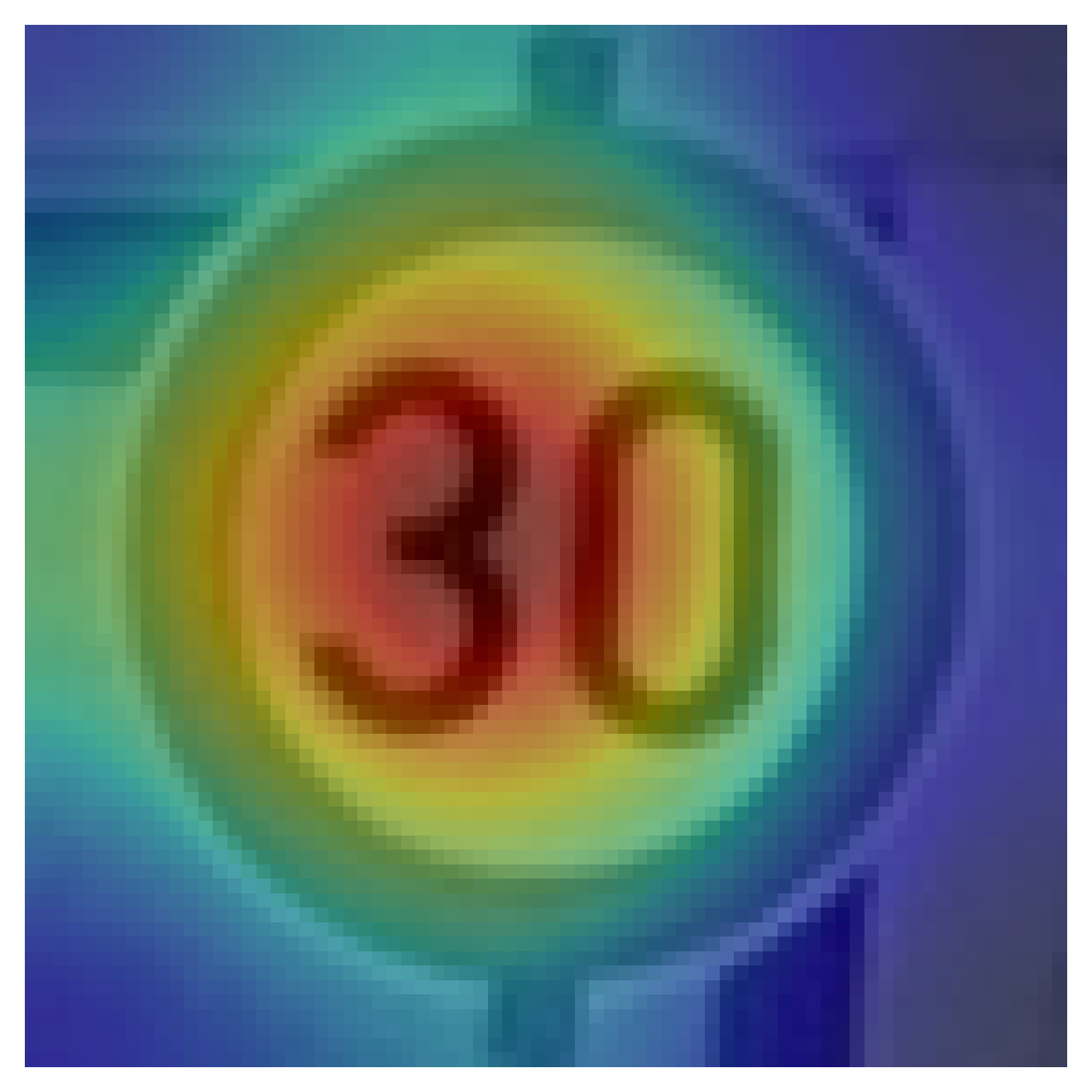}
         \caption{Heatmaps for the smaller \povd{} for ResNet-50. From left to right\begin{enumerate*}[before=\unskip{: }, itemjoin={{; }}, itemjoin*={{, and }}, label={(\roman*)}]
            \item 10cm, full circle
            \item 10cm, two sectors
            \item 20cm, full circle
            \item 20cm, two sectors.
         \end{enumerate*}}
         \label{fig:heatmaps_speedlimit_small}
     \end{subfigure}
     \caption{Heatmaps for the stop sign (upper row), and the 30km/h speed limit sign (middle row) for two different shapes, and two \ac{ML} models each. The lower row shows the heat maps for two different \povd{} sizes with ResNet-50 \ac{GTSRB}.}
     \label{fig:heatmaps}
\end{figure}

\begin{figure}[t]
     \centering
     \begin{subfigure}[b]{0.98\linewidth}
         \centering
         \includegraphics[width=0.24\textwidth]{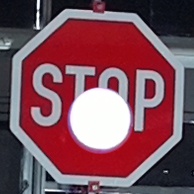}
         \includegraphics[width=0.24\textwidth]{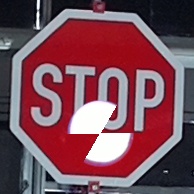}
         \includegraphics[width=0.24\textwidth]{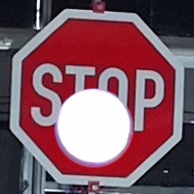}
         \includegraphics[width=0.24\textwidth]{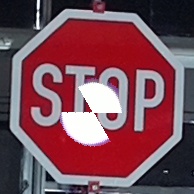}
         \caption{Examples for the stop sign. From left to right\begin{enumerate*}[before=\unskip{: }, itemjoin={{; }}, itemjoin*={{, and }}, label={(\roman*)}]
            \item ResNet-50, full circle
            \item ResNet-50, two sectors
            \item ConvNeXt small, full circle
            \item ConvNeXt small, two sectors.
         \end{enumerate*}}
         \label{fig:digital_simulation_stop}
     \end{subfigure}
     \hfill
     \begin{subfigure}[b]{0.98\linewidth}
         \centering
         \includegraphics[width=0.24\textwidth]{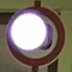}
         \includegraphics[width=0.24\textwidth]{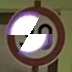}
         \includegraphics[width=0.24\textwidth]{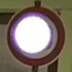}
         \includegraphics[width=0.24\textwidth]{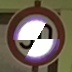}
         \caption{Examples for the 30km/h speed limit sign. From left to right\begin{enumerate*}[before=\unskip{: }, itemjoin={{; }}, itemjoin*={{, and }}, label={(\roman*)}]
            \item ResNet-50, full circle
            \item ResNet-50, two sectors
            \item ConvNeXt small, full circle
            \item ConvNeXt small, two sectors.
         \end{enumerate*}}
         \label{fig:digital_simulation_speedlimit}
     \end{subfigure}
     \hfill
     \begin{subfigure}[b]{0.98\linewidth}
         \centering
         \includegraphics[width=0.24\textwidth]{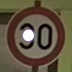}
         \includegraphics[width=0.24\textwidth]{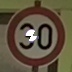}
         \includegraphics[width=0.24\textwidth]{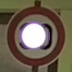}
         \includegraphics[width=0.24\textwidth]{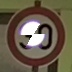}
         \caption{Examples for smaller \povd{}. From left to right\begin{enumerate*}[before=\unskip{: }, itemjoin={{; }}, itemjoin*={{, and }}, label={(\roman*)}]
            \item 10cm, full circle
            \item 10cm, two sectors
            \item 20cm, full circle
            \item 20cm, two sectors.
         \end{enumerate*}}
         \label{fig:digital_simulation_speedlimit_small}
     \end{subfigure}
     \caption{Examples of our digitally simulated \povd{}. The positions are derived from the heatmaps of Figure~\ref{fig:heatmaps}.}
     \label{fig:digital_simulation}
\end{figure}

We run the digital simulation of our \povd{} attack and obtain four heatmaps from the two shapes and the two \ac{ML} models, with one heatmap for each shape–model combination, as shown in Figure~\ref{fig:heatmaps_stop}. Although all heatmaps are qualitatively similar, ConvNeXt small \ac{GTSRB} shows a more focused area than ResNet-50 \ac{GTSRB}. A depiction of the two selected shapes for the highest-ranked positions in both \ac{ML} models is shown in Figure~\ref{fig:digital_simulation_stop}. \autoref{tab:digitalSimulationlabelDistribution} shows the top three labels of misclassified images from the digital simulation.

\begin{table}[t]
\centering
\caption{Top three labels of misclassified, digitally simulated images for the 30cm \povd{} for a stop sign and 30km/h speed limit sign. For ConvNeXt small at the stop sign, a maximum of two labels is observed.}
\label{tab:digitalSimulationlabelDistribution}
\resizebox{\linewidth}{!}{%
    \begin{tabular}{m{0.06\linewidth}|C{0.25\linewidth}L{0.3\linewidth}L{0.3\linewidth}}\Xhline{0.8pt}
         \textbf{Sign} & \multicolumn{1}{c}{\textbf{Model}} & \textbf{Full Circle} \fullcircle{0.8em} & \textbf{Two Sectors} \twoslices{0.5em} \\\Xhline{0.8pt}
         \multirow{5}{*}{\includegraphics[width=0.6cm]{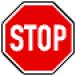}} & \multirow{2}{*}{\textbf{ResNet-50}} & Road Work & Road Work  \\
         & \multirow{2}{*}{\textbf{\ac{GTSRB}}} & No Passing & No Passing  \\
          & & No Vehicles & Pedestrians  \\\cline{2-4}
         & \textbf{ConvNeXt} & No Vehicles & No Passing End  \\
         & \textbf{small \ac{GTSRB}} & & Trucks Prohibited \\\hline
         \multirow{6}{*}{\includegraphics[width=0.6cm]{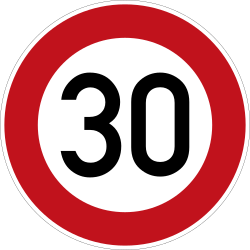}} & \multirow{2}{*}{\textbf{ResNet-50}} & No Vehicles & Speed Limit 50  \\
         & \multirow{2}{*}{\textbf{\ac{GTSRB}}} & Speed Limit 50 & No Passing  \\
          & & Speed Limit 80 & Priority Road  \\\cline{2-4}
         & \multirow{2}{*}{\textbf{ConvNeXt}} & No Vehicles & No Passing End  \\
         & \multirow{2}{*}{\textbf{small \ac{GTSRB}}} & No Passing End & No Vehicles \\
          & & Speed Limit 70 & No Passing  \\\Xhline{0.8pt}
    \end{tabular}}
\end{table}

\subsubsection{30km/h Speed Limit Sign}
Similarly, we run the digital simulation for a German 30km/h speed limit sign with the same shapes and the same \ac{ML} models with Figure~\ref{fig:heatmaps_speedlimit} depicting the resulting heatmaps and Figure~\ref{fig:digital_simulation_speedlimit} showing the digitally simulated \povd{} on the highest-ranked positions. In comparison to the stop sign, the heatmaps of the different shapes show a higher similarity, while the different \ac{ML} models show a clear difference in their position. Similar to our previous test with the stop sign, \autoref{tab:digitalSimulationlabelDistribution} contains the top misclassifications. Especially misclassifications of a different speed limit can lead to potentially safety-critical behaviour in real-world scenarios.

\subsubsection{30km/h Speed Limit Sign with smaller \povd{}}
As previously defined, we also run the digital simulation with 10cm and 20cm \povd{}s. 
We run the digital simulation for the same shapes as in the previous tests, resulting in the heatmaps shown in Figure~\ref{fig:heatmaps_speedlimit_small}. For all smaller \povd{} sizes that are shown in Figure~\ref{fig:digital_simulation_speedlimit_small}, this results in the highest number of misclassifications as a "Speed Limit 50." This digital simulation of smaller \povd{}s shows that even more stealthy versions of our attack are possible and cause misclassifications.

\subsection{Physical Test}
To evaluate the attack requirements defined in our threat model, we organize our results into categories targeting distinct attack capabilities. In particular, by comparing classification confidence with the \povd{} switched on and off while remaining physically deployed, we validate the triggerability requirement and demonstrate that the attack can be externally activated or deactivated without redeployment, enabling selective activation against specific targets.

\subsubsection{Stop Sign}
\begin{figure}[t]
     \centering
     \begin{subfigure}[b]{0.49\linewidth}
         \centering
         \includegraphics[width=0.48\textwidth]{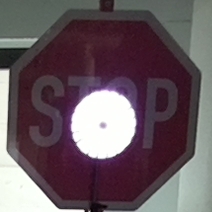}
         \includegraphics[width=0.48\textwidth]{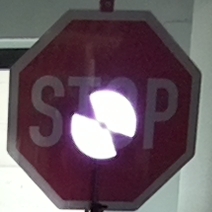}
         \caption{Example images in front of a stop sign}
         \label{fig:physical_copy_stop}
     \end{subfigure}
     \hfill
     \begin{subfigure}[b]{0.49\linewidth}
         \centering
         \includegraphics[width=0.48\textwidth]{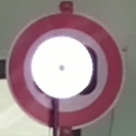}
         \includegraphics[width=0.48\textwidth]{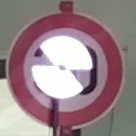}
         \caption{Example images in front of a 30km/h speed limit sign}
         \label{fig:physical_copy_speedlimit}
     \end{subfigure}
     \caption{Example images of the 30cm \povd{} for the physical tests. The positions are derived from \autoref{fig:heatmaps}.}
     \label{fig:physical_copy}
\end{figure}

In our first test, we place the \povd{} in front of a real stop sign at the aggregated position derived from the heatmaps of Figure~\ref{fig:heatmaps_stop}. With this test, we aim for a physical reproducibility of the results from Figure~\ref{fig:digital_simulation_stop} and \autoref{tab:digitalSimulationlabelDistribution}. We capture images with a Sony IMX708 from distances of 5m to 20m to investigate the effect of different distances. Example images, captured at a distance of 5m, are available in Figure~\ref{fig:physical_copy_stop}.

As our analysis of the \ac{ASR} in \autoref{tab:physical_copy_ASR} shows, the attack is successful for both ResNet-50 \ac{GTSRB} and ConvNeXt small \ac{GTSRB}, with a minimum \ac{ASR} of 40.69\% for ConvNeXt small, and 99.70\% for ResNet-50. While ResNet-50 shows a high \ac{ASR} across all distances and shapes, ConvNeXt small shows greater success at longer distances. The \acp{ASR} confirm the placement provided by our digital simulation, but the top misclassification labels differ, as they are not targeted by the simulation. A major reason is the differently perceived red color of the benign stop sign image in the digital simulation compared to the actual color from physical measurements, which are affected by reflections and variations in white balancing.

To ensure that the misclassifications are not the result of either the physical mounting of our \povd{} or of the differently perceived colors, we perform measurements for all distances with switched-off \acp{LED} of the \povd{} but the same mounting and illumination conditions as in \autoref{fig:physical_copy}. ResNet-50 \ac{GTSRB} classifies all images correctly as a stop sign with an average confidence of 97.90\%, and ConvNeXt small \ac{GTSRB} classifies them all correctly with an average confidence of 71.49\%. This test also shows that \povd{} is not harmful unless the \acp{LED} are triggered by the attacker.

\begin{table}[t]
\centering
\caption{Overview of the \ac{ASR} and top misclassification label for the physical tests of the 30cm \povd{} in front of different traffic signs.}
\label{tab:physical_copy_ASR}
\resizebox{\linewidth}{!}{%
\includegraphics[width=\linewidth]{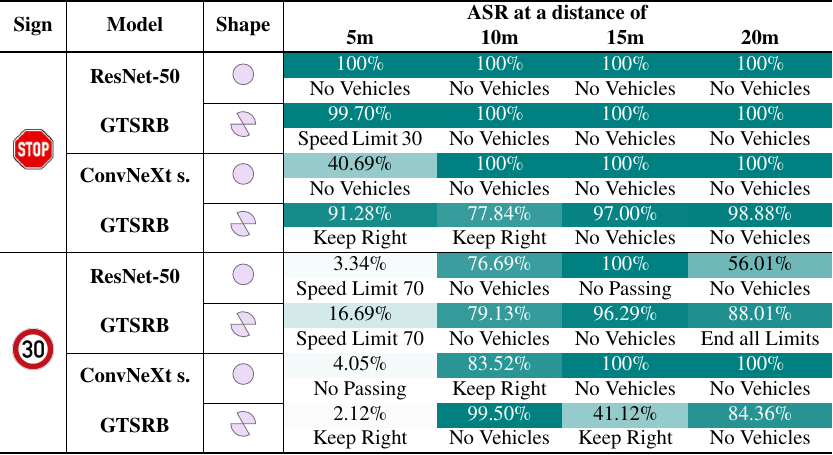}
}
\end{table}

\subsubsection{30km/h Speed Limit Sign}
Similar to the stop sign, we evaluate the \povd{} in front of the 30km/h speed limit sign for distances of 5m to 20m and use the heatmaps of Figure~\ref{fig:heatmaps_speedlimit} for the placement. Since the heatmaps for ResNet-50 \ac{GTSRB} and ConvNeXt small \ac{GTSRB} show different central points, we capture all images for two placement options. Figure~\ref{fig:physical_copy_speedlimit} shows the physical reproducibility of one exemplary position to reproduce the digital simulations of Figure~\ref{fig:digital_simulation_speedlimit}, captured at a distance of 5m.

The \ac{ASR} is high for most test cases with the 30km/h speed limit sign, as shown in \autoref{tab:physical_copy_ASR}, but especially ConvNeXt small shows low \acp{ASR} for the 5m distance, while it is more successful at distances of 10m to 20m. The misclassification labels from digital simulation in \autoref{tab:digitalSimulationlabelDistribution} show similarities, especially for the "No Vehicles" misclassification. At the same time, there are similar optical artifacts in the real-world deployment as for the stop sign.

For the speed limit sign, both models classify all images for all distances as a 30km/h speed limit sign if the rotating \povd{} has the \acp{LED} switched off. While ResNet-50 \ac{GTSRB} has an average confidence of 99.95\%, ConvNeXt small \ac{GTSRB} shows 67.24\%.

\subsubsection{30km/h Speed Limit Sign with smaller \povd}
\begin{figure}[t]
     \centering
     \begin{subfigure}[b]{0.32\linewidth}
         \centering
         \includegraphics[height=0.455\textwidth]{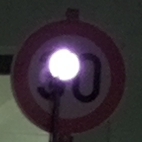}
         \includegraphics[height=0.455\textwidth]{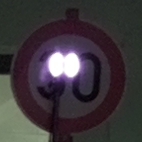}
         \caption{10cm\,\povd}
        \vspace{0.85\baselineskip}
         \label{fig:physical_copy_speedlimit_10cm}
     \end{subfigure}
     \hfill
     \begin{subfigure}[b]{0.32\linewidth}
         \centering
         \includegraphics[height=0.445\textwidth]{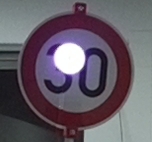}
         \includegraphics[height=0.445\textwidth]{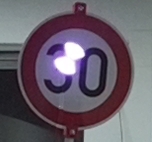}
         \caption{15cm portable \povd}
         \label{fig:physical_copy_speedlimit_15cm}
     \end{subfigure}
     \hfill
     \begin{subfigure}[b]{0.32\linewidth}
         \centering
         \includegraphics[height=0.455\textwidth]{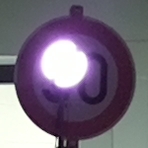}
         \includegraphics[height=0.455\textwidth]{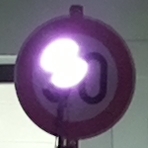}
         \caption{20cm\,\povd}
        \vspace{0.85\baselineskip}
         \label{fig:physical_copy_speedlimit_20cm}
     \end{subfigure}
     \caption{Example images of smaller \povd{}s for the physical tests, each with two different shapes. The positions are derived from Figure~\ref{fig:heatmaps_speedlimit_small}. The differently perceived brightness levels result from the exposure control algorithm of the camera to compensate for the increased brightness of the larger \povd.}
     \label{fig:physical_copy_small}
\end{figure}

For the smaller 10cm, 15cm, and 20cm \povd, we evaluated the same distances from 5m to 20m as for the other physical tests, based on the heatmaps of Figure~\ref{fig:heatmaps_speedlimit_small}. Figure~\ref{fig:physical_copy_small} shows example images of the smaller \povd{}s, displaying both a full circle and two sectors.

\begin{table}[t]
\centering
\caption{Overview of the \ac{ASR} and top misclassification label for the physical tests of the small \povd{}s in front of the 30km/h speed limit sign.}
\label{tab:physical_copy_ASR_small}
\resizebox{\linewidth}{!}{%
\includegraphics[width=\linewidth]{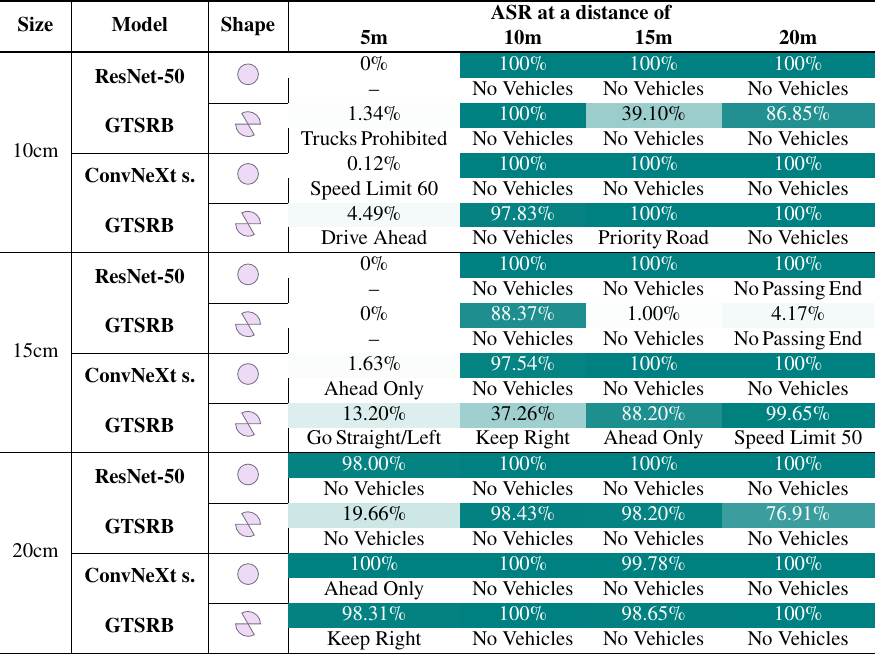}
}
\end{table}

As shown in \autoref{tab:physical_copy_ASR_small}, the 20cm \povd{} achieves high ASRs across all tested distances, whereas the smaller 10cm \povd{} becomes similarly effective only for distances of at least 10m.
Nevertheless, misclassification persists up to 10m, meaning that correct recognition may occur only very late, leaving limited reaction time for perception systems. The portable 15cm \povd{} is effective in most cases, except for ResNet-50 \ac{GTSRB} and two circular sectors. The lower \ac{ASR} at a distance of 5m is similar to the observation of our 10cm \povd{}. When comparing the top misclassification labels with those from the larger 30cm \povd{} in \autoref{tab:physical_copy_ASR}, they show high similarity for most test cases. This evaluation shows that even a smaller, stealthier \povd{} attack can be effectively deployed in the real world.

Similar to our other physical tests, both ResNet-50 \ac{GTSRB} and ConvNeXt small \ac{GTSRB} classify all street signs with \acp{LED} switched of but rotating \povd{} correctly, except for a small number of misclassifications for ConvNeXt small at 5m and 10m. While ResNet-50 \ac{GTSRB} has an average confidence of 99,94\% for the switched-off 10cm \povd{} and 99.97\% for the 20cm version, ConvNeXt small \ac{GTSRB} shows 74.93\% and 75.67\%, respectively.

\subsubsection{Impact of Illumination}
As the \povd{} needs to fulfill the previously described timing constraints of \autoref{eq:povConstraint}, tests at bright daylight are not possible, as the short exposure times of cameras would require an extremely fast rotating engine. We will, therefore, conduct night tests, where the timing constraint can be fulfilled. For this evaluation, we create a test stand consisting of two automotive \ac{LED} car headlights~\cite{osram_gmbh_night_2025} that we align as they are used in an average compact car. Additionally, we calibrate the headlights using UN Regulation No. 112~\cite{united_nations_eece324rev2add111rev4_2023}. This calibration is important to create realistic conditions, especially considering the retroreflective property of street signs~\cite{din_deutsches_institut_fur_normung_e_v_din_2025, european_committee_for_standardization_en_2007, united_nations_economic_commission_for_europe_ecetrans196_2006}. Similar to other tests, we repeat our evaluation at distances of 5m to 20m.

\begin{table}[t]
\centering
\caption{Overview of the \ac{ASR} for the physical night tests of the 30cm \povd{} in front of different traffic signs.}
\label{tab:night_ASR}
\resizebox{\linewidth}{!}{%
\includegraphics[width=\linewidth]{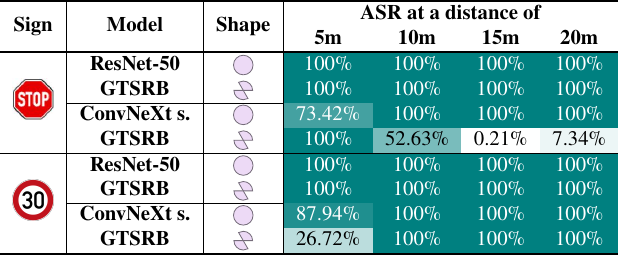}
}
\end{table}

\begin{figure}[t]
     \centering
     \begin{subfigure}[b]{0.25\linewidth}
         \centering
         \includegraphics[width=1\textwidth]{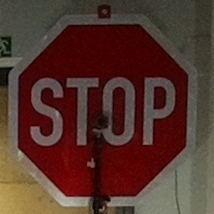}
         \caption{Switched off}
         \label{fig:night_tests_benign}
     \end{subfigure}
     \hfill
     \begin{subfigure}[b]{0.25\linewidth}
         \centering
         \includegraphics[width=1\textwidth]{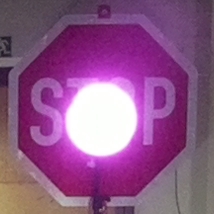}
         \caption{Full circle}
         \label{fig:night_tests_fullcircle}
     \end{subfigure}
     \hfill
     \begin{subfigure}[b]{0.25\linewidth}
         \centering
         \includegraphics[width=1\textwidth]{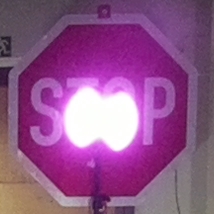}
         \caption{Two sectors}
         \label{fig:night_tests_twohalves}
     \end{subfigure}
     \caption{Example images of our \povd{} in front of a stop sign for evaluating the impact of illumination. The effect of the near-infrared \acp{LED} is more noticeable, as the camera adjusts its exposure settings to the darker environment.}
     \label{fig:night_tests}
\end{figure}

As shown in \autoref{tab:night_ASR}, our \povd{} shows even higher \acp{ASR} for close distances at all models, compared to the results of \autoref{tab:physical_copy_ASR}. The only exceptions are the two sectors at distances of 15m and 20m and ConvNeXt small \ac{GTSRB}. In these scenarios, the two sectors cannot be captured accurately at night, as the timing requirement of \autoref{eq:povConstraint} is not fulfilled. This leads to imprecise shapes. For closer distances, the attack shows even higher \acp{ASR}, as the near-infrared \acp{LED} are more noticeable. \autoref{fig:night_tests} shows that the glare effect is stronger compared to the indoor day tests of \autoref{fig:physical_copy}. For both the stop sign and the 30km/h speed limit sign, we used the 30cm version of our \povd. If the \povd{} is not triggered, all images of both signs are correctly classified.

\subsubsection{Impact of Placement}
\begin{figure}[t]
\centering
     \begin{subfigure}[b]{0.25\linewidth}
         \centering
         \includegraphics[width=1\textwidth]{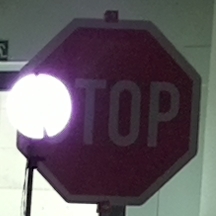}
         \caption{Full circle}
         \label{fig:stop_sign_off_position_fullcircle}
     \end{subfigure}
     \hspace{0.5cm}
     \begin{subfigure}[b]{0.25\linewidth}
         \centering
         \includegraphics[width=1\textwidth]{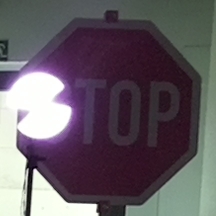}
         \caption{Two sectors}
         \label{fig:stop_sign_off_position_twohalves}
     \end{subfigure}
    \caption{Example images of the \povd{} arbitrarily placed outside the heatmaps of Figure~\ref{fig:heatmaps_stop}.}
    \label{fig:stop_sign_off_position}
\end{figure}

To demonstrate that effective \povd{} deployment is highly placement-dependent, we evaluate an intentionally non-optimal placement of the \povd{} in front of a stop sign. Specifically, we place a 30 cm \povd{} at a position that lies outside the high-impact regions identified by our digital simulation heatmaps. This experiment shows that placing the \povd{} at arbitrary locations does not reliably cause misclassification, highlighting the necessity of our digital simulation for identifying effective placement locations. As in previous experiments, we evaluate both the full circle and the two circular sectors at distances ranging from 5m to 20m using ResNet-50 and ConvNeXt-Small trained on \ac{GTSRB}. The \povd{} is positioned to partially cover the letter “S” of the stop sign, as shown in \autoref{fig:stop_sign_off_position}. 

\begin{table}[t]
\centering
\caption{Overview of the \ac{ASR} for the \povd{} placed outside the heatmaps positions, in front of a stop sign. The \ac{ASR} is significantly lower compared to the recommended placement at the heatmap positions in \autoref{tab:physical_copy_ASR}.}
\label{tab:stop_sign_off_position}
\resizebox{\linewidth}{!}{%
\includegraphics[width=\linewidth]{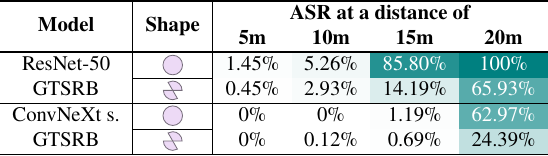}
}
\end{table}

\autoref{tab:stop_sign_off_position} shows that the \ac{ASR} is substantially lower compared to the correctly placed \povd{} in \autoref{tab:physical_copy_ASR}. In particular, for shorter and intermediate distances, the off-position deployment fails to cause misclassification for ConvNeXt-Small \ac{GTSRB} and has only a negligible effect on ResNet-50 \ac{GTSRB}. While a higher ASR can be observed at larger distances, the attack effectiveness remains inconsistent and significantly lower than that achieved with heatmap-guided placement. Overall, these results demonstrate that arbitrary \povd{} placement does not reliably induce misclassification and highlight the importance of our digital simulation in identifying effective deployment locations.

Similar to our previous physical tests, we also perform tests at this position with a switched-off \povd{} to evaluate the impact of the mounting and the physical deployment of the \povd. All captured images are classified correctly by both \ac{ML} models for all distances, with an average confidence of 99.93\% for ResNet-50 \ac{GTSRB} and 75.95\% for ConvNeXt small \ac{GTSRB}. These values are similar to the correctly placed and switched-off \povd{}, highlighting that the misclassifications in \autoref{tab:stop_sign_off_position} are solely from the \povd{} with activated \ac{LED} and not from other influences.

\subsubsection{Attack Transferability}
To evaluate the transferability of our \povd{} attack across different perception models, we test the same physically captured attack images on multiple \ac{ML} models. We reuse the identical physical test scenarios from our initial physical tests: a 30cm \povd{} placed in front of a stop sign and a 30km/h speed limit sign, with distances ranging from 5m to 20m, and evaluate the models listed in \autoref{tab:mlModelsTransferability}. We analyze the transferability of the speed limit sign only for \ac{GTSRB}-trained models, as only these models are trained to classify this traffic sign. To isolate transferability from deployment effects, we fix the placement using the positions from the heatmaps in \autoref{fig:heatmaps}, ensuring that differences in attack success are from the model behavior rather than the attack procedure or the digital simulation. 

\begin{table}[t]
\centering
\caption{The \ac{ASR} of the transferabillity tests of the 30cm \povd{} in front of different traffic signs.}
\label{tab:transferability_30cm}
\resizebox{\linewidth}{!}{%
\includegraphics[width=\linewidth]{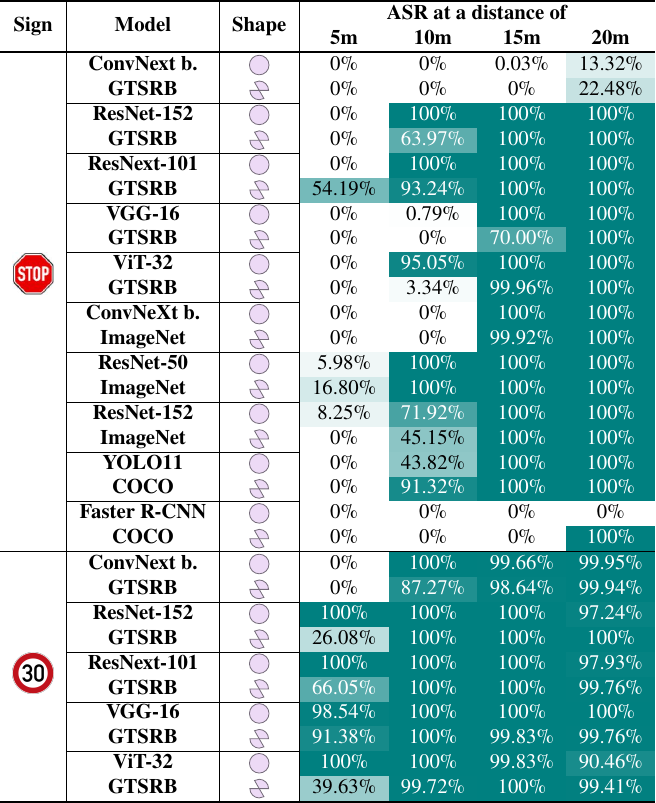}
}
\end{table}

\autoref{tab:transferability_30cm} shows the results of our transferability analysis of the 30cm \povd{}. Although close distances of 5m show no attack success for most models with a stop sign, the attack remains effective against the majority of models and distances of $\geq$10m. For the 30km/h speed limit sign, the attack remains effective against most models even at close range. The \ac{ML} model performance of all tested use-cases of \autoref{tab:transferability_30cm} shows no, or a negligible number of misclassifications if the \povd{} is rotating but switched off, except for ResNet-50 ImageNet at 15m and 20m.

\begin{table}[t]
\centering
\caption{The \ac{ASR} of the transferability tests of the 10, 15, and 20cm \povd{} in front of the 30km/h speed limit sign.}
\label{tab:transferability_small}
\resizebox{\linewidth}{!}{%
\includegraphics[width=\linewidth]{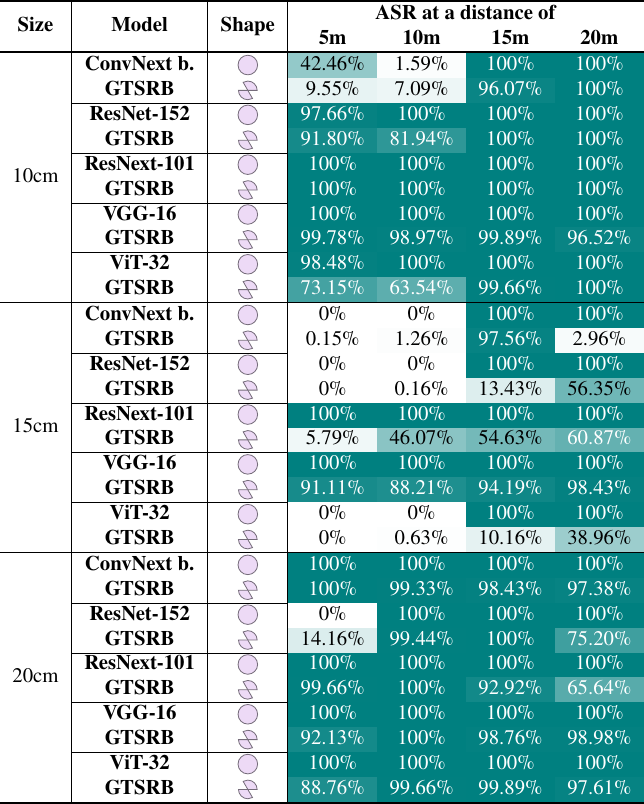}
}
\end{table}

We also test the transferability of the smaller 10cm, 15cm, and 20cm \povd{} in front of the 30km/h speed limit sign. Similarly, we test distances from 5m to 20m on all \ac{GTSRB} models specified in \autoref{tab:mlModelsTransferability}. Like the speed limit sign results of the 30cm \povd{} in \autoref{tab:transferability_30cm}, the smaller \povd{}s in \autoref{tab:transferability_small} show high transferability for the majority of tested cases. The \ac{ML} models perform with no or a negligible amount of misclassifications if the smaller \povd{}s are switched off, except for ViT-32 \ac{GTSRB} at 5m and 20m. The transferability results show that near-infrared \povd{}s can pose a severe threat for many \ac{ML} models, independent of their training dataset or model architecture.

\subsection{Dynamic Test}

\begin{figure}[t]
\centering
     \begin{subfigure}[b]{0.3\linewidth}
         \centering
         \includegraphics[width=1\textwidth]{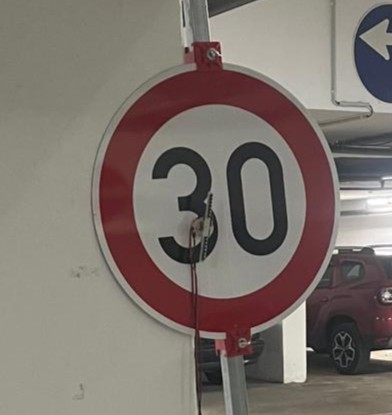}
         \caption{\povd{} deployment}
         \label{fig:dynamic_Test_deployment}
     \end{subfigure}
     \hspace{0.5cm}
     \begin{subfigure}[b]{0.58\linewidth}
         \centering
         \includegraphics[width=1\textwidth]{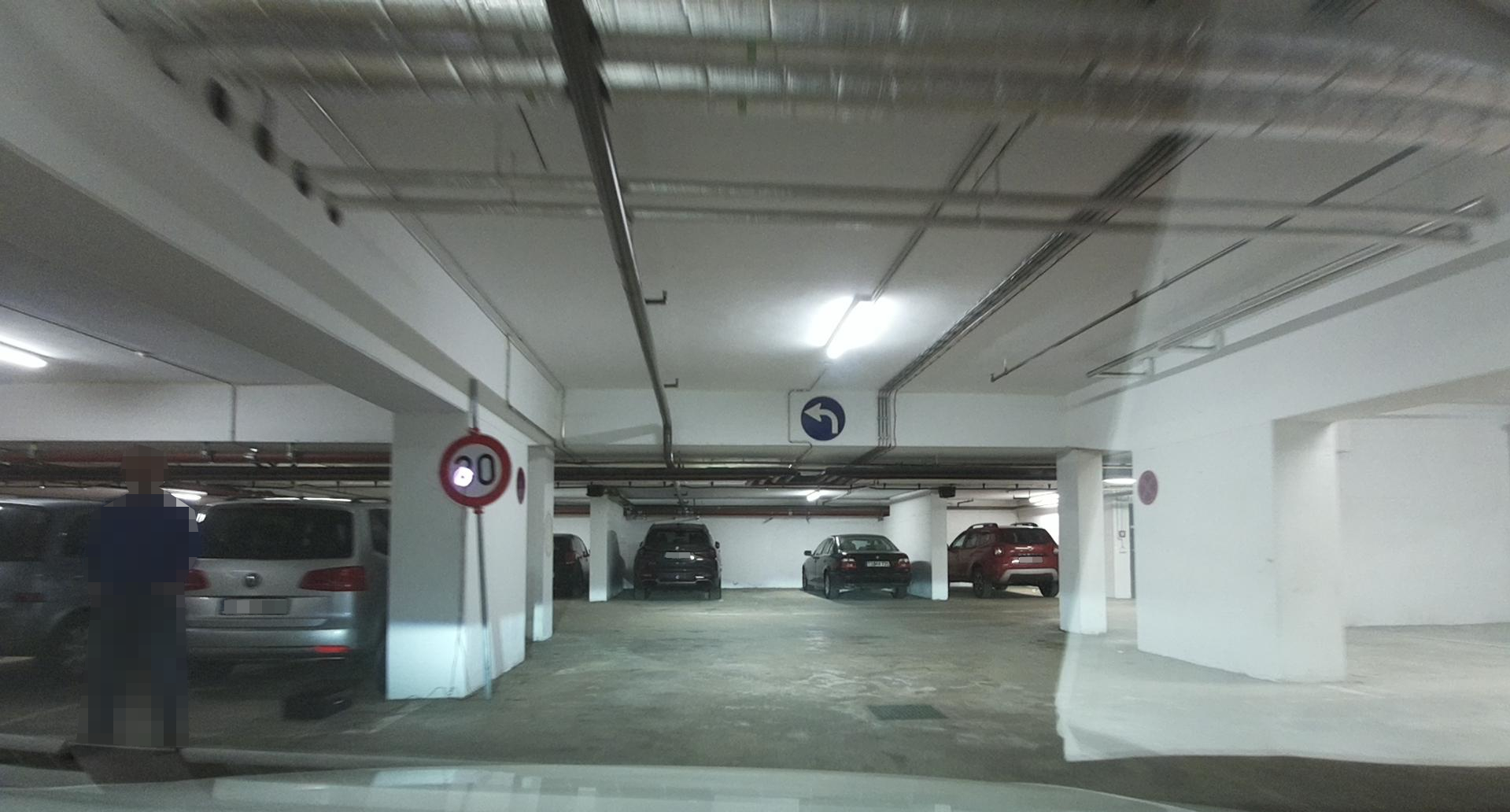}
         \caption{Camera perspective of the \povd{}}
         \label{fig:dynamic_Test_camera_perspective}
     \end{subfigure}
    \caption{Test setup of the dynamic tests. The portable \povd{} is attached to the traffic sign via magnets and perceived by the camera as a bright circle.}
    \label{fig:dynamic_Test}
\end{figure}

In this evaluation, we aim to investigate the impact of driving dynamics on the \ac{ASR}. We place our camera, the Sony IMX708 with no infrared filter~\cite{raspberry_pi_ltd_camera_2024}, in a real vehicle and place the 30km/h speed limit sign with the 15cm portable \povd{} in an indoor parking lot. For our tests, we start at a distance of 30m from the traffic sign and approach it at a maximum speed of 10km/h to stay legally compliant. The \povd{} deployment and the camera perspective of our dynamic tests are shown in \autoref{fig:dynamic_Test}. Similar to our static physical tests, we capture videos and evaluate the \ac{ASR} as the ratio of misclassified frames per test while passing by the traffic sign. To ensure repeatability of our results, we conduct five tests with this setup.
\begin{table}[t]
\centering
\caption{Overview of \acp{ASR} of the dynamic real-world tests. Even in dynamic environments, the \povd{} remains effective against most \ac{ML} models. As with other evaluations using the speed limit sign, all models are used in their \ac{GTSRB} variants.}
\label{tab:dynamicTests}
\resizebox{1\linewidth}{!}{%
    \includegraphics[width=\linewidth]{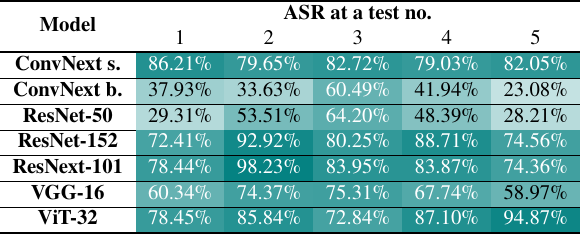}}
\end{table}
As shown in \autoref{tab:dynamicTests}, the \povd{} is also effective in dynamic environments, with high \acp{ASR} for most evaluated \ac{ML} models. For the tested models, ConvNext base \ac{GTSRB} shows the lowest \ac{ASR} across all tests, which is consistent with the transferability analysis of the 15cm \povd{} in \autoref{tab:transferability_small}. Similarly, ResNet-50 \ac{GTSRB} shows lower \acp{ASR} in all tests. This is the same as the lower \acp{ASR} for the 15cm \povd{} in \autoref{tab:physical_copy_ASR_small}. The other models show an \ac{ASR} of $\geq$58\% in all our tests.
\section{Defense}

In this section, we discuss a hardware-based defense mechanism using an optical filter and propose a software-based detection method.

\subsubsection{Hardware-based Defense}
\begin{figure}[t]
     \centering
     \begin{subfigure}[b]{0.28\linewidth}
         \centering
         \includegraphics[width=1\textwidth]{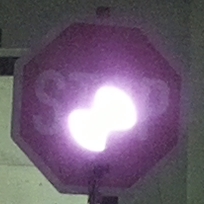}
         \caption{\textit{Without}\,near-infrared\,cutoff\,filter}
         \label{fig:defense_ir}
     \end{subfigure}
     \hspace{0.5cm}
     \begin{subfigure}[b]{0.28\linewidth}
         \centering
         \includegraphics[width=1\textwidth]{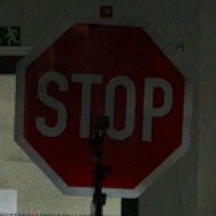}
         \caption{\textit{With}\,near-infrared\,cutoff\,filter}
         \label{fig:defense_rgb}
     \end{subfigure}
     \caption{Comparison of a \povd{} showing two sectors, captured with the same cameras, only adding a near-infrared cutoff filter. Although the identical image sensor is used, the infrared \povd{} is not visible anymore.}
     \label{fig:defense}
\end{figure}

As our proposed \povd{} operates in the near-infrared spectrum, the most effective defense is an infrared cutoff filter. Such optical filters will block light of the near-infrared spectrum and are available with different cutoff wavelengths~\cite{bte_bedampfungstechnik_gmbh_infrared-filter_2025, optics_balzers_ag_coated_2025}. We evaluate the impact of a near-infrared cutoff filter using the same image sensor, a Sony IMX708~\cite{raspberry_pi_ltd_camera_2024}, with and without the filter. This ensures that only the impact of the filter is evaluated, not other sensor parameters or data processing. As shown in \autoref{fig:defense}, the impact of a near-infrared cutoff filter is significant: While the image sensor \textit{without} the filter shows the displayed content as in our previous evaluation, the same image sensor \textit{with} a filter does not perceive the \povd{} anymore. Although Figure~\ref{fig:defense_rgb} appears visually darker, it is not a direct effect of the cutoff filter but rather the absence of additional infrared reflections that are visible in Figure~\ref{fig:defense_ir}, when no filter is applied. While the images of the camera \textit{without} an infrared filter show an \ac{ASR} of 100\% for both ResNet-50 \ac{GTSRB} and ConvNeXt small \ac{GTSRB}, all images from the camera \textit{with} a filter are classified correctly with an average confidence of $\approx$93\%. We repeat the tests with a \povd{} that shows a full circle and obtain the same result. This \ac{ML} model-agnostic defense can also be effective against other infrared-based attacks~\cite{sato_invisible_2024, wang_i_2021}. While highly effective, such hardware modifications can be infeasible in already deployed autonomous vehicles. In addition, using this cutoff filter may conflict with design goals, such as low-light perception~\cite{sato_invisible_2024}. Adding an infrared filter can reduce image quality or sensing reliability in these cases. As a result, this defense may not be suitable for all camera setups or operating conditions.

\begin{figure}[t]
	\centering
	\includegraphics[width=0.95\linewidth]{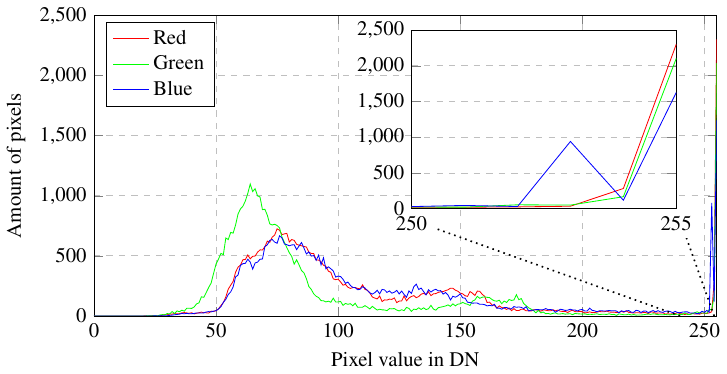}
	\caption{Histogram of Figure~\ref{fig:defense_ir}. For values >250, the number of red and blue pixels is higher than the green pixels.}
	\label{fig:defense_histogram}
\end{figure}

\subsubsection{Software-based Detection}
We discuss a sensor-specific, software-based detection approach that can be used to detect near-infrared \povd{} attacks. Due to differences in spectral sensitivity across image sensors, near-infrared light can be perceived as visible colors such as red, purple, or magenta~\cite{sato_invisible_2024}.
For the image sensor used in our evaluation, the Sony IMX708, the perceived color is magenta, as shown in the evaluation images. When creating a histogram of the attacked image in Figure~\ref{fig:defense_ir}, there is not only a highly saturated region with all color channels being at their maximum value, as shown in \autoref{fig:defense_histogram}. Due to spectral sensitivity, the close proximity of the saturated region results in a higher number of red and blue pixel values exceeding 250 than in the green channel. This anomaly is specific to the sensor hardware used and needs to be derived from the spectral response characteristics of the sensor.

Based on that observation, we provide a proof-of-concept detection approach in Algorithm~\ref{img:irPovdDetectionAlgo}.
Our approach first identifies the saturated regions in an image and creates (line 3) binary masks of these regions using a \ac{BFS}~\cite{cormen_elementary_2009}. For each saturated region, it analyzes the spatial proximity (lines 9-12) for the identified anomaly, specifically a higher occurrence of high red and blue pixel values compared to green ones (lines 19-21).
We evaluate this detection approach on more than 180k images from the test cases described in Section~\ref{sec:evaluation} and report the \ac{TP} and \ac{TN} rates. The \ac{TP} rate is the amount of correctly identified \povd{}s, while the \ac{TN} rate is important for the benign cases.
As shown in \autoref{tab:detectionEvaluation}, our approach achieves high \ac{TN} rates, with only two false positives among $\approx$27k benign images. While the \ac{TP} rate decreases for smaller \povd{}, it remains above 55\%. Overall, this software-based detection serves as a proof-of-concept that relies on sensor-specific artifacts. 
 
\begin{algorithm}[t]
\caption{Proposed detection of a near-infrared \povd{} in a captured image $I$.}
\label{img:irPovdDetectionAlgo}
\includegraphics[width=\linewidth]{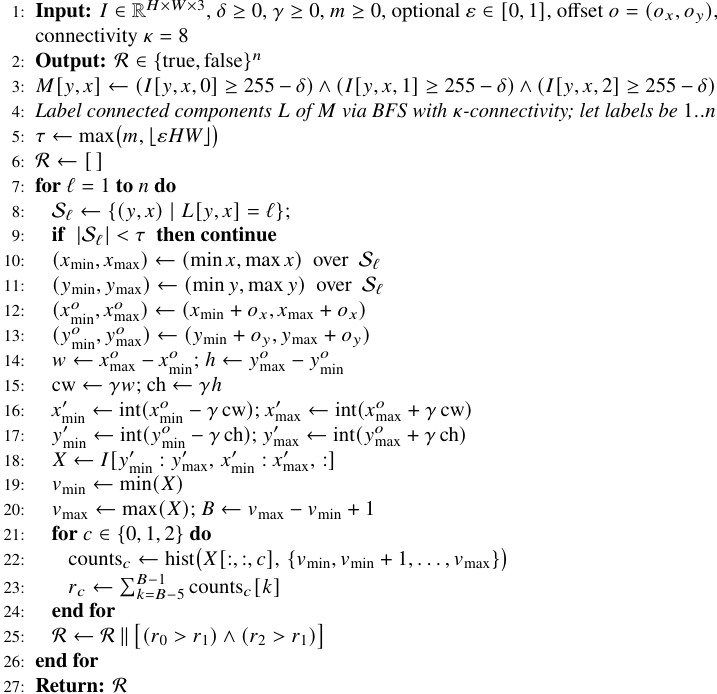}
\end{algorithm}

\begin{table}[t]
\centering
\caption{\ac{TN} and \ac{TP} rates for our proposed near-infrared \povd{} detection algorithm for over 180k images.}
\label{tab:detectionEvaluation}
\resizebox{1\linewidth}{!}{%
\begin{tabular}{L{0.55\linewidth}C{0.15\linewidth}C{0.15\linewidth}}\Xhline{0.8pt}
    \textbf{Test case} & \textbf{\ac{TN} rate} & \textbf{\ac{TP} rate} \\\Xhline{0.8pt}
    Stop Sign & 99.97\% & 79.25\% \\
    30km/h Sign & 100\% & 78.39\% \\
    30km/h Sign, 10cm \povd & 100\% & 66.26\% \\
    30km/h Sign, 20cm \povd & 100\% & 55.45\% \\
    Portable \povd & 100\% & 98.30\% \\
    Stop Sign Night & 100\% & 98.89\% \\
    30km/h Sign Night & 100\% & 89.08\% \\
    Different Placement & \textsuperscript{*} & 68.51\% \\\Xhline{0.8pt}
    \multicolumn{3}{l}{* Only adversarial images available in this use case.}
\end{tabular}
}
\end{table}

\section{Discussion}
\label{sec:discussion}
Our evaluation results show that near-infrared \povd{}s can pose a significant threat to traffic sign classification models, leading to misclassifications and transferring well to other \ac{ML} models. We further discuss two practical limitations of near-infrared \povd{} attacks:

\begin{enumerate}[nosep]
   \item The effectiveness of near-infrared \povd{}s depends on the ability of infrared light to reach the image sensor and is therefore limited to infrared-sensitive cameras. 
   \item The perceived color of near-infrared emissions depends on sensor spectral sensitivity and may appear red, purple, or magenta~\cite{sato_invisible_2024}. As a result, near-infrared \povd{}s offer limited control over color and primarily allow shape-based attack patterns.
\end{enumerate}

To explore whether these limitations are specific to near-infrared operation rather than inherent to the \povd{} concept, we additionally investigate \povd{}s operating in the human-visible spectrum using RGB \acp{LED}.
Compared with our attack goals in Section~\ref{sec:attackGoalRequirements}, RGB \povd{}s sacrifice parts of the stealthiness, as they operate in the human-visible spectrum and can be perceived by humans. Nevertheless, RGB \povd{}s remain less conspicuous than classical displays used in prior attacks~\cite{chahe_dynamic_2024, patel_overriding_2022}. While both classical displays and \povd{}s can be remotely triggered to show dynamic attack content, \povd{}s appear visually transparent when rotating without displaying content, as human vision cannot follow the fast rotation speed. This makes \povd{}s stealthier than classical displays, which are typically black and framed when switched off.

In contrast to near-infrared \povd{}s, RGB \povd{}s can not only visualize different shapes but also controllable colors that are perceived by cameras using the widely available Bayer color filter array~\cite{bayer_color_1976}. This additional degree of freedom enables the visualization of dynamic content and colors while preserving the attack goals described in Section~\ref{sec:attackGoalRequirements}.

\begin{table}[t]
\centering
\caption{Overview of the \acp{ASR} for different colored circles of the RGB \povd{} in front of a stop sign.}
\label{tab:appRGBEvaluation}
\resizebox{\linewidth}{!}{%
\includegraphics[width=\linewidth]{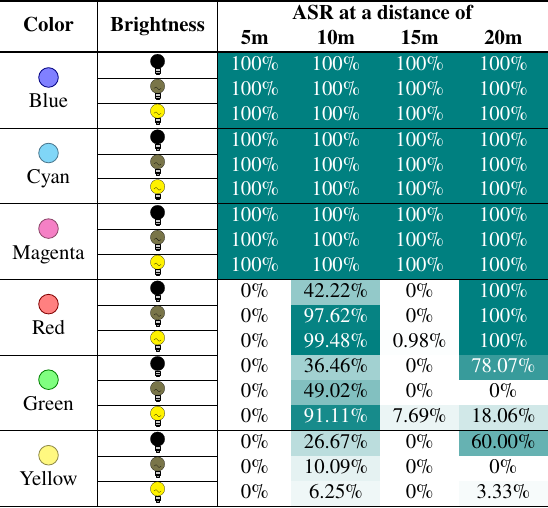}
}
\end{table}

\begin{figure}[t]
     \centering
     \begin{subfigure}[b]{0.32\linewidth}
         \centering
         \includegraphics[width=0.8\textwidth]{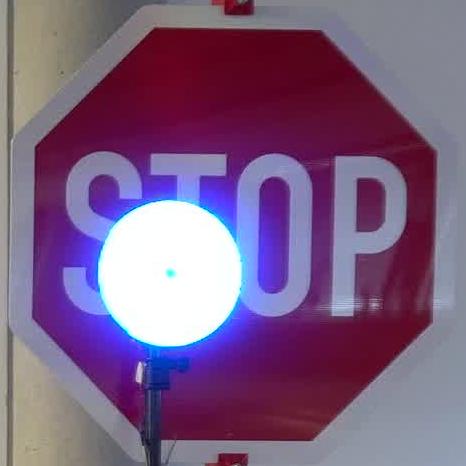}
         \caption{Blue~(\raisebox{-0.5ex}{\fullcircleRGB{0.8em}{blue}})}
         \label{fig:appRGBBlue}
     \end{subfigure}
     \hfill
     \begin{subfigure}[b]{0.32\linewidth}
         \centering
         \includegraphics[width=0.8\textwidth]{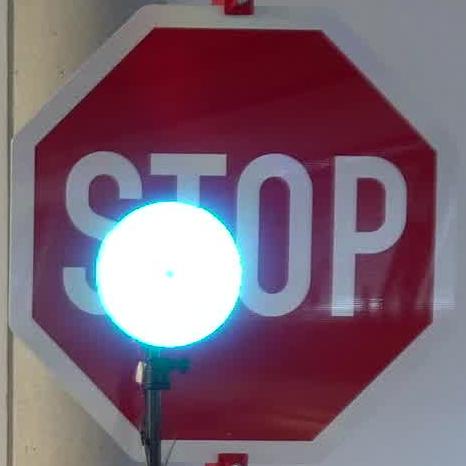}
         \caption{Cyan~(\raisebox{-0.5ex}{\fullcircleRGB{0.8em}{cyan}})}
         \label{fig:appRGBCyan}
     \end{subfigure}
     \hfill
     \begin{subfigure}[b]{0.32\linewidth}
         \centering
         \includegraphics[width=0.8\textwidth]{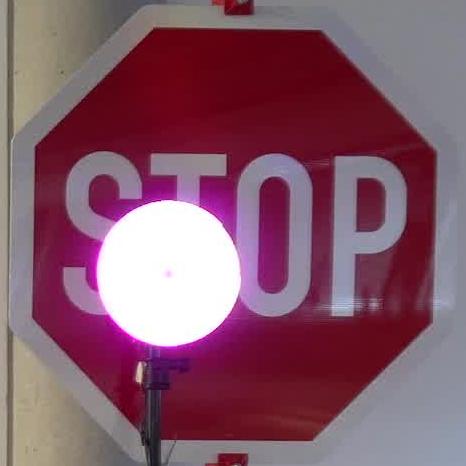}
         \caption{Magenta~(\raisebox{-0.5ex}{\fullcircleRGB{0.8em}{magenta}})}
         \label{fig:appRGBMagenta}
     \end{subfigure}
     \hfill
     \begin{subfigure}[b]{0.32\linewidth}
         \centering
         \includegraphics[width=0.8\textwidth]{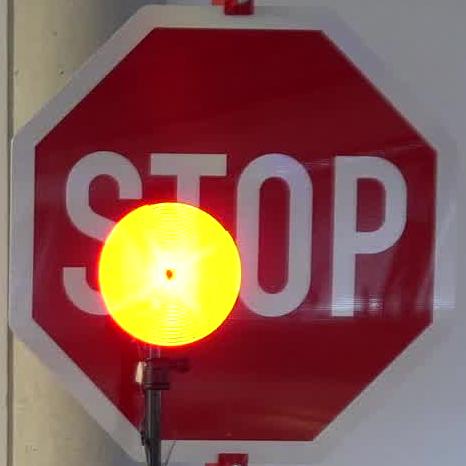}
         \caption{Red~(\raisebox{-0.5ex}{\fullcircleRGB{0.8em}{red}})}
         \label{fig:appRGBRed}
     \end{subfigure}
     \hfill
     \begin{subfigure}[b]{0.32\linewidth}
         \centering
         \includegraphics[width=0.8\textwidth]{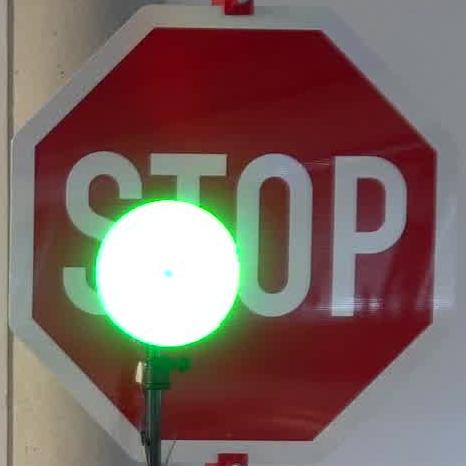}
         \caption{Green~(\raisebox{-0.5ex}{\fullcircleRGB{0.8em}{green}})}
         \label{fig:appRGBGreen}
     \end{subfigure}
     \hfill
     \begin{subfigure}[b]{0.32\linewidth}
         \centering
         \includegraphics[width=0.8\textwidth]{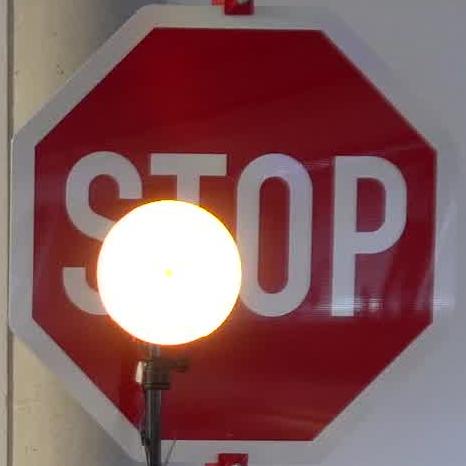}
         \caption{Yellow~(\raisebox{-0.5ex}{\fullcircleRGB{0.8em}{yellow}})}
         \label{fig:appRGBYellow}
     \end{subfigure}
     \caption{Example images from the RGB \povd{} with different-colored full circles. The images have been captured at a distance of 5m with the \povd{} at full brightness.}
     \label{fig:appRGB}
\end{figure}

We conduct an initial evaluation of an RGB \povd{} prototype by capturing videos using a Sony ILCE-6400~\cite{sony_electronics_inc_ilce-6400_2025} and the same definition of the \ac{ASR} as in our previous evaluation. We test the following colors: Blue~(\raisebox{-0.5ex}{\fullcircleRGB{0.8em}{blue}}), Cyan~(\raisebox{-0.5ex}{\fullcircleRGB{0.8em}{cyan}}), magenta~(\raisebox{-0.5ex}{\fullcircleRGB{0.8em}{magenta}}), red~(\raisebox{-0.5ex}{\fullcircleRGB{0.8em}{red}}), green~(\raisebox{-0.5ex}{\fullcircleRGB{0.8em}{green}}), and yellow~(\raisebox{-0.5ex}{\fullcircleRGB{0.8em}{yellow}}). Additionally, we we set the RGB \acp{LED} to three different brightness levels: Bright~(\raisebox{-0.4ex}{\resizebox{0.7em}{!}{\lightbulb{yellow}}}; 100\% intensity), medium~(\raisebox{-0.4ex}{\resizebox{0.7em}{!}{\lightbulb{yellow!40!black}}}; 60\% intensity), and dark~(\raisebox{-0.4ex}{\resizebox{0.7em}{!}{\lightbulb{black}}}; 20\% intensity). Following our evaluation from Section~\ref{sec:evaluation}, we capture 30-second video snippets at distances from 5m to 20m and use the same \ac{ASR} definition as in our previous experiments. For this evaluation, we execute all steps of the attack design shown in \autoref{fig:attack_overview} using a stop sign and targeting ResNet-50 \ac{GTSRB}. As \autoref{tab:appRGBEvaluation} shows, the RGB \povd{} is especially effective when displaying blue, cyan, or magenta circles, while red, green, and yellow show a lower \ac{ASR}. Example images captured at 5m are shown in \autoref{fig:appRGB}.

\section{Conclusion}
Our near-infrared \povd{} represents a new physical adversarial attack, operating in a light spectrum outside the human-visible spectrum with high rotation speeds to ensure stealthiness. 
Unlike prior infrared-based attacks~\cite{sato_invisible_2024, wang_i_2021}, \povd{}s support dynamic and remotely triggerable attack content, enabling flexible and targeted deployment. Through extensive evaluation across different traffic signs, \ac{ML} model architectures, distances, and environmental conditions, we demonstrated that \povd{}s can reliably cause misclassifications and exhibit strong transferability across \ac{ML} models. Our results further show that correct placement, as determined through a digital simulation, is important for attack success. We also demonstrated that smaller, portable \povd{}s remain effective, increasing the practicality of real-world attacks.

Additionally, we discussed the impact of hardware- and sensor-dependent factors on near-infrared \povd{}s and showed that \povd{} attacks exhibit a balance between stealthiness and deployability in camera-based perception systems. To explore the design space further, we introduced an RGB-based \povd{} variant that trades partial stealth for increased control over displayed content, highlighting that the \povd{} concept extends beyond a single wavelength spectrum. Overall, our findings demonstrate that our proposed attack represents a realistic and flexible threat to camera-based perception systems.

\bibliographystyle{IEEEtran}
\bibliography{bibliography}

\end{document}